\documentclass[superscriptaddress,onecolumn,showpacs,
amssymb,amsmath,nobibnotes,aps,prd,
nofootinbib]{revtex4}
\pdfoutput=1
\usepackage{graphicx,subfigure,bm,color,psfrag,hyperref}
\usepackage{amsfonts}
\usepackage{lipsum}
\usepackage{mathtools}
\usepackage{verbatim}
\usepackage{color}
\begin{document}

\title{Observational constraints on one-parameter dynamical dark-energy parametrizations and the $H_0$ 
tension}

\author{Weiqiang Yang}
\email{d11102004@163.com}
\affiliation{Department of Physics, Liaoning Normal University, Dalian, 116029, P. R. 
China}

\author{Supriya Pan}
\email{supriya.maths@presiuniv.ac.in}
\affiliation{Department of Mathematics, Presidency University, 86/1 College Street, 
Kolkata 700073, 
India}

\author{Eleonora Di Valentino}
\email{eleonora.divalentino@manchester.ac.uk}
\affiliation{Jodrell Bank Center for Astrophysics, School of Physics and Astronomy, 
University of 
Manchester, Oxford Road, Manchester, M13 9PL, UK}

\author{Emmanuel N. Saridakis}
\email{Emmanuel\_Saridakis@baylor.edu}
\affiliation{Physics Division, National Technical University of Athens, 15780 Zografou 
Campus,
Athens, Greece}
\affiliation{CASPER, Physics Department, Baylor University, Waco, TX 76798-7310, USA}

\author{Subenoy Chakraborty}
\email{schakraborty@math.jdvu.ac.in}
\affiliation{Department of Mathematics, Jadavpur University, Kolkata 700032, West Bengal, 
India}

\begin{abstract}
The phenomenological parametrizations of dark-energy (DE) equation of state can be very 
helpful, since they allow for the investigation of its cosmological behavior despite the 
fact that its underlying theory is unknown.  However, although   there has been a large amount of research on DE parametrizations which involve 
two or more free parameters, the one-parameter parametrizations seem to be 
underestimated. We perform a detailed observational confrontation of 
five one-parameter DE models, with  observational data from   
cosmic microwave background (CMB), Joint light-curve analysis sample 
from Supernovae Type Ia observations (JLA),  baryon acoustic oscillations  (BAO) 
distance measurements, and cosmic chronometers (CC). We find that all models favor a 
phantom DE equation of state at present time, while they lead to  $H_0$ values
 in perfect agreement with its direct measurements and therefore they offer an 
alleviation to the $H_0$-tension. Finally, performing a Bayesian analysis we show that 
although 
$\Lambda$CDM cosmology is still favored, one-parameter DE models have similar or better 
efficiency in fitting the data comparing to   two-parameter DE parametrizations, and 
thus they deserve a thorough investigation.

\end{abstract}


\maketitle
 
\section{Introduction}

The remarkable journey of modern cosmology started in 1998, when the observational 
evidences showed that  we are living in an accelerating universe and that the previous
physical scenarios  needed to be retraced. The introduction of dark energy 
(DE) concept was  a need in order for the observational predictions to acquire a 
solid theoretical formulation. The dark energy is  a component with high negative pressure 
that drives the universe acceleration, nevertheless its nature has remained a mysterious chapter in the scientific history after a series of investigations by a 
large number of researchers. The cosmological constant is the simplest DE fluid 
with the above features, however the ``cosmological constant 
problem'' \cite{Weinberg:1988cp} and the possibility that the DE sector could be dynamical 
led to a number of explanations, mainly in two directions. The first is to consider 
that the DE sector corresponds to a peculiar extra fluid that fills the universe in the 
framework of general relativity \cite{Copeland:2006wr,Cai:2009zp}. The second direction 
is to consider that the DE fluid is an effective one, arising from a modification of the 
gravitational sector itself   \cite{modgrav1,Capozziello:2011et,Cai:2015emx}. 

Independently of the underlying nature and the micro-physical theory of DE, one can 
introduce  phenomenological parametrizations of the DE equation-of-state parameter $w_x = 
p_x/\rho_x$, where $p_x$ and $\rho_x$ are respectively the pressure and energy density of 
the (effective) DE perfect fluid, which is considered to have a dynamical character in 
general. Since for the moment we do not have any fundamental 
rule in favor of some specific equation-of-state parameters, we may 
consider various functional forms for $w_x$. For a literature survey of various DE 
parametrizations and models we refer to the works   \cite{Chevallier:2000qy, Linder:2002et, 
Cooray:1999da, Efstathiou:1999tm, 
Astier:2000as, Weller:2001gf, Jassal:2005qc, Linder:2005ne,Gong:2005de,
Nesseris:2005ur, Feng:2004ff, Xia:2006rr, Basilakos:2006us,
Nojiri:2006ww,Saridakis:2008fy,Barboza:2008rh,Saridakis:2009pj,Dutta:2009yb,
Saridakis:2009ej, Ma:2011nc,Feng:2011zzo, Feng:2012gf, 
DeFelice:2012vd, Chen:2011cy,Basilakos:2013vya,Umilta:2015cta,Ballardini:2016cvy,DiValentino:2016hlg, Chavez:2016epc,
DiValentino:2017zyq,DiValentino:2017gzb,Zhao:2017cud,DiValentino:2017rcr,Yang:2017amu, 
Marcondes:2017vjw,Yang:2017alx, Pan:2017zoh,Vagnozzi:2018jhn}. 

In general, the well known DE  parametrizations have two free parameters, usually denoted 
by $w_0w_a$CDM models, where $w_0$ marks the present value of $w_x$ and $w_a$ 
characterizes the dynamical nature of the DE sector. However, apart from the $w_0w_a$CDM 
parametrizations, one-parameter dynamical DE parametrizations, as well as models with 
more than two parameters, have also been introduced and investigated in the last 
years. 
Nevertheless, the one-parameter dynamical DE parametrizations are much neglected in the 
literature compared to the DE parametrizations having two or more parameters. In 
principle we do not find any strong reason behind this underestimation, and thus in this work we 
aim to investigate the features of this particular class of DE 
parametrizations, and explore its cosmological viabilities with the recent observational 
evidences, taking into account their advantage that they are more economical and have less number of 
free parameters compared to other dark energy models with two or more than two free parameters.    

Hence,  we introduce various one-parameter dynamical DE parametrizations that 
are primarily motivated from the phenomenological ground, and we perform a detailed 
observational confrontation. In particular, we use data from   
cosmic microwave background (CMB) observations, from Joint light-curve analysis sample 
from Supernovae Type Ia observations (JLA), from baryon acoustic oscillations  (BAO) 
distance measurements, as well as from cosmic chronometers  Hubble parameter measurements 
(CC), performing additionally various combined analyses.

The manuscript is organized as follows. In Section \ref{sec-2} we present the basic 
equations for a general dark-energy scenario at both the background and perturbation
level, and we display the five one-parameter DE parametrizations that are going to be 
investigated. In Section \ref{sec-data} we describe the observational data sets that  
will be used. In Section \ref{sec-results} we perform the observational confrontation, 
extracting the observational constraints on the various cosmological quantities. After that in Section \ref{sec-baysian} we compare the present dynamical DE parametrizations mainly through the Bayesian analysis. Finally,  we close the present work in Section \ref{sec-conclu} with a brief summary.

\section{One-parameter parametrizations  at background and perturbation levels}
\label{sec-2}

In this section we present the basic equations that determine the universe evolution 
at both the background and perturbation levels, and we  introduce various 
one-parameter parametrizations for the dark-energy equation-of-state parameter. 
Throughout the work we consider the homogeneous and isotropic  
Friedmann-Lema\^{i}tre-Robertson-Walker (FLRW) geometry, with metric
\begin{eqnarray}
{\rm d}s^2 = -{\rm d}t^2 + a^2 (t) \left[\frac{{\rm d}r^2}{1-Kr^2} + r^2 \left(d \theta^2 
+ \sin^2 \theta d \phi^2\right)\right],
\end{eqnarray}
where $a(t)$  is the scale factor  and $K$ determines the spatial curvature, with values 
$0$, $-1$ and $+1$ corresponding to  spatially flat, open and closed universe, respectively. 

We consider a universe filled with radiation, baryons and cold dark matter, and we additionally consider the  DE fluid. In this case the Friedmann equations, that determine the 
universe evolution at the background level, read as
\begin{eqnarray}
H^2 + \frac{K}{a^2} &=& \frac{8\pi G}{3} \rho_{tot},\label{efe1}\\
2\dot{H} + 3 H^2  + \frac{K}{a^2} &=& - 8 \pi G\, p_{tot}\label{efe2},
\end{eqnarray}
with $G$ the Newton's constant and $H=\dot{a}/a$ the Hubble function of the FLRW universe, with dots 
denoting derivatives with respect to cosmic time. In the above expressions we have 
introduced the total energy density and pressure as $\rho_{tot} = \rho_r +\rho_b +\rho_c 
+\rho_x$ and $p_{tot} = p_r + 
p_b + p_c + p_x$ respectively, with the symbols $r,\; b,\; c,\; x$ denoting 
radiation, baryon, cold dark matter and   dark energy fluids.
Finally, for simplicity in the following we will focus on the spatially flat case  
($K=0$) since it is favored by observations.

As usual  we assume that the above sectors  do not have any mutual interaction, 
and thus  the conservation equation of each fluid   is
\begin{eqnarray}\label{cons}
\dot{\rho}_i + 3 H (1 +w_i ) \rho_i = 0,
\end{eqnarray} 
where $i \in \{ r, b, c, x\}$ and $p_i = w_i \rho_i$, $w_i$ being the barotropic state parameter for the $i$-th fluid. 
Note that out of equations  (\ref{efe1}), (\ref{efe2}) and (\ref{cons}), only two are 
independent. Hence, using the known equation-of-state parameters  $w_r=1/3$, $w_b=w_c=0$, in  (\ref{cons})  one can explicitly write down the 
conservation equations for radiation, baryons and cold dark matter respectively as, $\rho_r =\rho_{r0}\, a^{-4}$, $\rho_b = 
\rho_{b0}
\, a^{-3}$ and $\rho_c = \rho_{c0}\, a^{-3}$, with $\rho_{i0}$   the present value of 
$\rho_i$ and where we have set the present scale factor   $a_0$ to $1$. Similarly, 
concerning the dark energy sector, equation (\ref{cons}) leads to
\begin{eqnarray}\label{de-evol}
\rho_{x}=\rho_{x,0}\,\left(  \frac{a}{a_{0}}\right)  ^{-3}\,\exp\left[
-3\int_{a_{0}}^{a}\frac{w_{x}\left(  a'\right)  }{a'}\,da'
\right].
\end{eqnarray}
 Thus, the evolution equation (\ref{de-evol}) implies that the dynamics of DE can 
be determined as long as a specific parametrization of the DE equation of state is given.

Having presented the equations that determine the universe evolution at the background 
level, we now proceed to the investigation of its evolution at the perturbation level, 
since this is related to the observed structure formation. In order to study the  
perturbation equations, one needs to consider the perturbed 
FLRW metric either in synchronous  or in conformal Newtonian gauge. In the following  
we consider the former choice, in which the perturbed metric takes the form 
\begin{eqnarray}
\label{perturbed-metric}
ds^2 = a^2(\eta) \left [-d\eta^2 + (\delta_{ij}+h_{ij}) dx^idx^j  \right],
\end{eqnarray}
where $\eta$ is the conformal time, and $\delta_{ij}$,  $h_{ij}$ are 
respectively the unperturbed and the perturbed metric tensors.  Now, in  the synchronous gauge the conservation equations of energy and momentum for the $i$-th component of the fluid for a mode with 
wavenumber ${k}$  can be written as \cite{Mukhanov,Ma:1995ey, Malik:2008im}:
\begin{eqnarray}
&&\delta'_{i}  = - (1+ w_{i})\, \left(\theta_{i}+ \frac{h'}{2}\right) - 
3\mathcal{H}\left(\frac{\delta p_i}{\delta \rho_i} - w_{i} \right)\delta_i - 9 
\mathcal{H}^2\left(\frac{\delta p_i}{\delta \rho_i} - c^2_{a,i} \right) (1+w_i) 
\frac{\theta_i}
{{k}^2}, \label{per1} \\
&&\theta'_{i}  = - \mathcal{H} \left(1- 3 \frac{\delta p_i}{\delta 
\rho_i}\right)\theta_{i} 
+ \frac{\delta p_i/\delta \rho_i}{1+w_{i}}\, {k}^2\, \delta_{i} 
-{k}^2\sigma_i,\label{per2}
\end{eqnarray}
where primes mark derivatives with respect to conformal time and  $\mathcal{H}= 
a^{\prime}/a$ is the conformal Hubble parameter. Furthermore,  $\delta_i = 
\delta \rho_i/\rho_i$ is the density perturbation for the $i$-th fluid,  
$\theta_{i}\equiv i k^{j} v_{j}$ denotes the divergence of the $i$-th fluid 
velocity, $h = h^{j}_{j}$ stands for the trace of the metric perturbations $h_{ij}$, and  
$\sigma_i$ is the anisotropic stress related to the $i$-th fluid. Finally, the quantity 
$c_{a,i}^2 = \dot{p}_i/\dot{\rho}_i$ denotes the adiabatic speed of sound of the $i$-th 
fluid, and it is given by $ c^2_{a,i} =  w_i - \frac{w_i^{\prime}}{3\mathcal{H}(1+w_i)}$
in the case where we set the sound  speed $c^2_{s} = \delta p_i / \delta \rho_i$ to $1$.
In the following analysis we neglect the anisotropic stress for simplicity.

\begin{figure}[ht]
    \centering
       \includegraphics[width=0.495\textwidth]{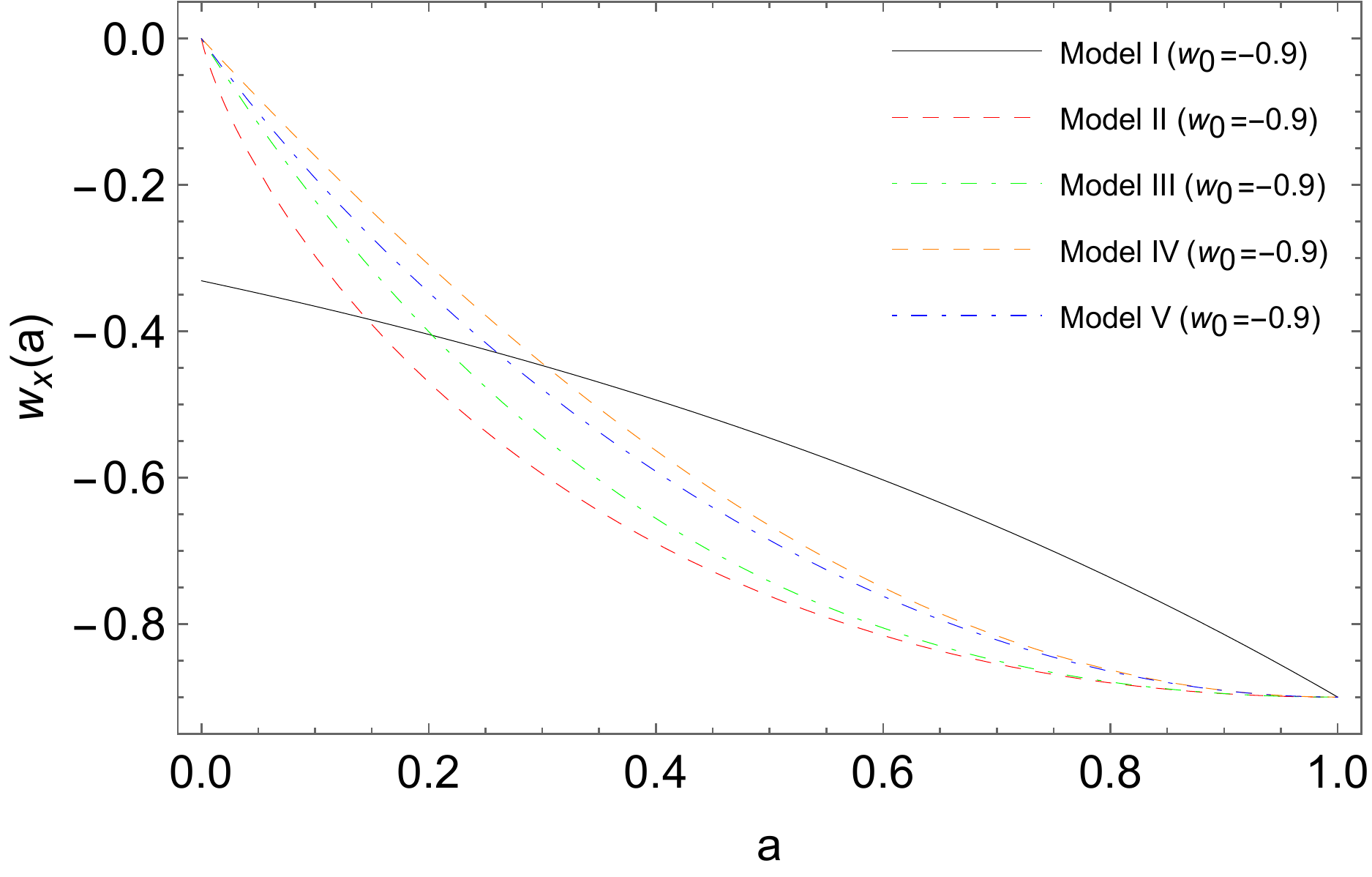}
    \includegraphics[width=0.495\textwidth]{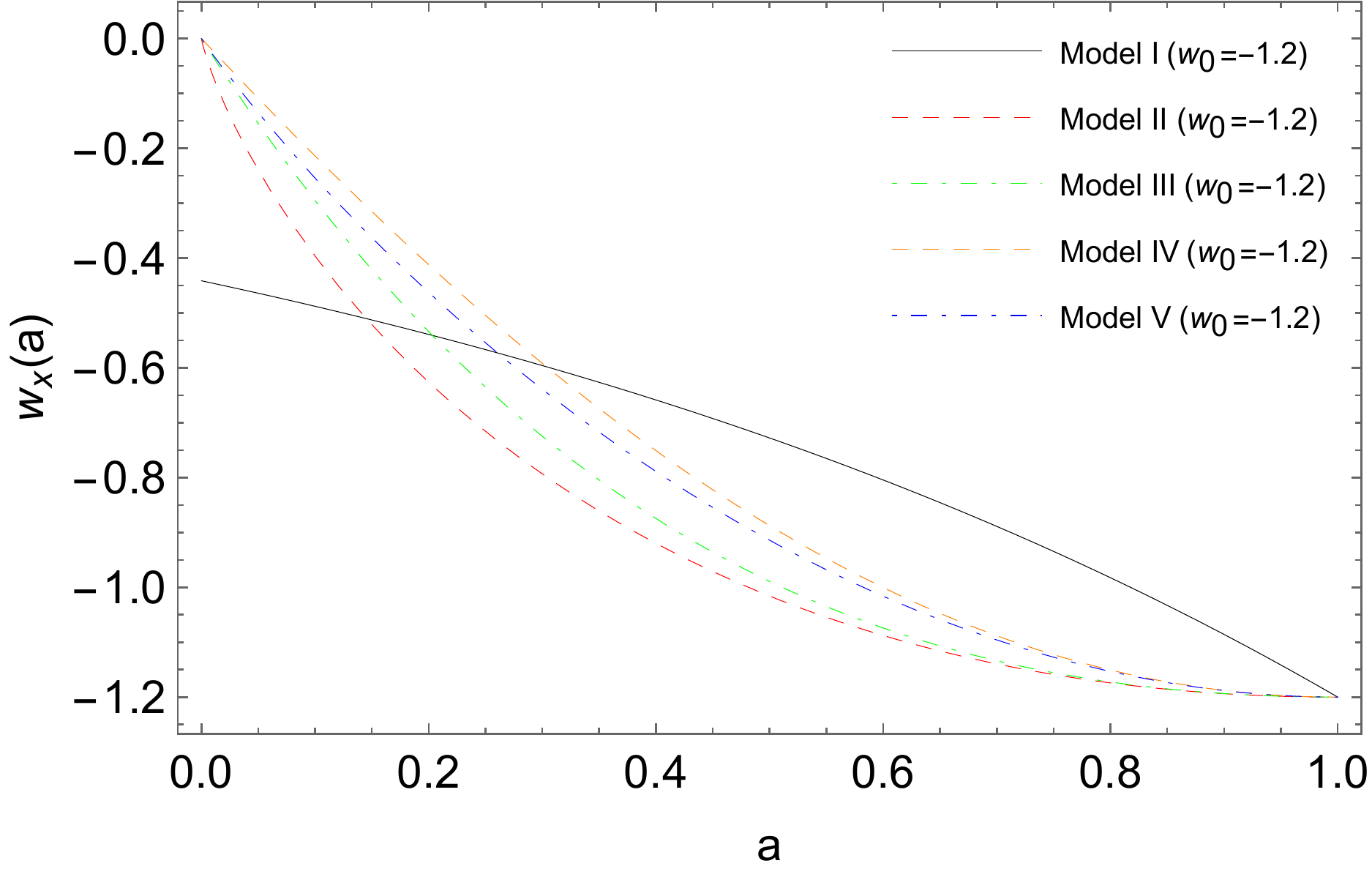}
     \caption{{\it{The evolution of the one-parameter dynamical DE equation-of-state 
parametrizations (\ref{model1})-(\ref{model5})  as a function of the scale factor, for  
 $w_0= -0.9$  (left graph) and  $w_0 = -1.2$
(right graph).}}}
\label{fig:wa}
\end{figure}

In this work we are interested in investigating one-parameter DE equation-of-state 
parametrizations. In particular, we consider five such parametrizations given by:
\begin{eqnarray}
&&{\rm Model~I}:~~~~w_x(a)=w_0\exp(a-1),\label{model1}\\
&&{\rm Model~II}:~~~~w_x(a)=w_0a[1-\log(a)],\label{model2}\\
&&{\rm Model~III}:~~~~w_x(a)=w_0a\exp(1-a),\label{model3}\\
&&{\rm Model~IV}:~~~~w_x(a)=w_0a[1+\sin(1-a)],\label{model4}\\
&&{\rm Model~V}:~~~~w_x(a)=w_0a[1+\arcsin(1-a)],\label{model5}
\end{eqnarray}
where $w_0$ is the only  free parameter, corresponding to the   dark energy 
equation-of-state parameter at present. In order to provide a more transparent picture 
of the behavior of the above parametrizations, in Fig.~\ref{fig:wa} we depict  $w_x(a)$, 
taking two cases for $w_0$, namely one lying in the quintessence and one lying in the 
phantom regime. As we can see, in all models  $w_x(a)$ presents a decreasing behavior.

\section{Observational data}
\label{sec-data}

In this section we proceed to a detailed observational confrontation of the one-parameter 
dynamical DE equation-of-state parametrizations (\ref{model1})-(\ref{model5}) presented 
in the previous section. We analyze several combinations of cosmological data, by 
considering the six cosmological parameters of the standard $\Lambda$CDM paradigm: the 
baryon and the cold dark matter energy densities 
$\Omega_{\rm b}h^2$ and $\Omega_{\rm c}h^2$, the ratio between the sound horizon and the 
angular diameter distance at decoupling $\Theta_{s}$, the reionization optical depth 
$\tau$, and the spectral index and the amplitude of the scalar primordial power spectrum 
$n_\mathrm{S}$ and $A_\mathrm{S}$. 
Moreover, for the various models  we add the free parameter $w_0$,  which parametrizes 
the DE evolution. All these $7$ free parameters are explored within the range of the 
conservative flat priors listed in Table~\ref{priors}. 
\begin{table}[ht]
\begin{center}
\begin{tabular}{|c|c|}
\hline
Parameter                    & prior \\
\hline
$\Omega_{\rm b} h^2$         & $[0.013,0.033]$ \\
$\Omega_{\rm c} h^2$       & $[0.001,0.99]$ \\
$\Theta_{\rm s}$             & $[0.5,10]$ \\
$\tau$                       & $[0.01,0.8]$ \\
$n_\mathrm{S}$               & $[0.7,1.3]$ \\
$logA$                       & $[1.7, 5.0]$ \\
$w_0$                        & $[-2,0]$  \\
\hline 
\end{tabular}
\end{center}
\caption{Summary of the flat priors on the cosmological parameters assumed in 
this work, for the different DE parametrizations (\ref{model1})-(\ref{model5}).}
\label{priors}
\end{table}

We derive the bounds on the cosmological parameters by analyzing the full range of the 
2015 Planck temperature and polarization   cosmic microwave background (CMB) angular 
power spectra, and we call this combination ``CMB'' ~\cite{Adam:2015rua, Aghanim:2015xee}.
Additionally, we consider the Joint light-curve analysis sample from Supernovae Type Ia 
and we refer to this dataset as ``JLA'' ~\cite{Betoule:2014frx}. Furthermore, we add the 
baryon acoustic oscillations  (BAO) distance measurements, and we call them ``BAO'' 
~\cite{Beutler:2011hx, Ross:2014qpa,Gil-Marin:2015nqa}.
Finally, we use the Hubble parameter measurements from the cosmic chronometers (CC) and 
we refer to them as ``CC'' ~\cite{Moresco:2016mzx}.

In order to analyze statistically the several combinations of datasets, exploring the 
different dynamical DE scenarios, we use our modified version of the publicly 
available Monte-Carlo Markov Chain package \texttt{Cosmomc} \cite{Lewis:2002ah}, 
including the support for the Planck data release 2015 Likelihood Code 
\cite{Aghanim:2015xee}\footnote{See \url{http://cosmologist.info/cosmomc/}}. This has a 
convergence diagnostic based on the Gelman and Rubin statistic and implements an 
efficient sampling of the posterior distribution using the fast/slow parameter 
de-correlations \cite{Lewis:2013hha}.

\section{Results}
\label{sec-results}

\begin{figure}[ht]
\includegraphics[width=0.95\textwidth]{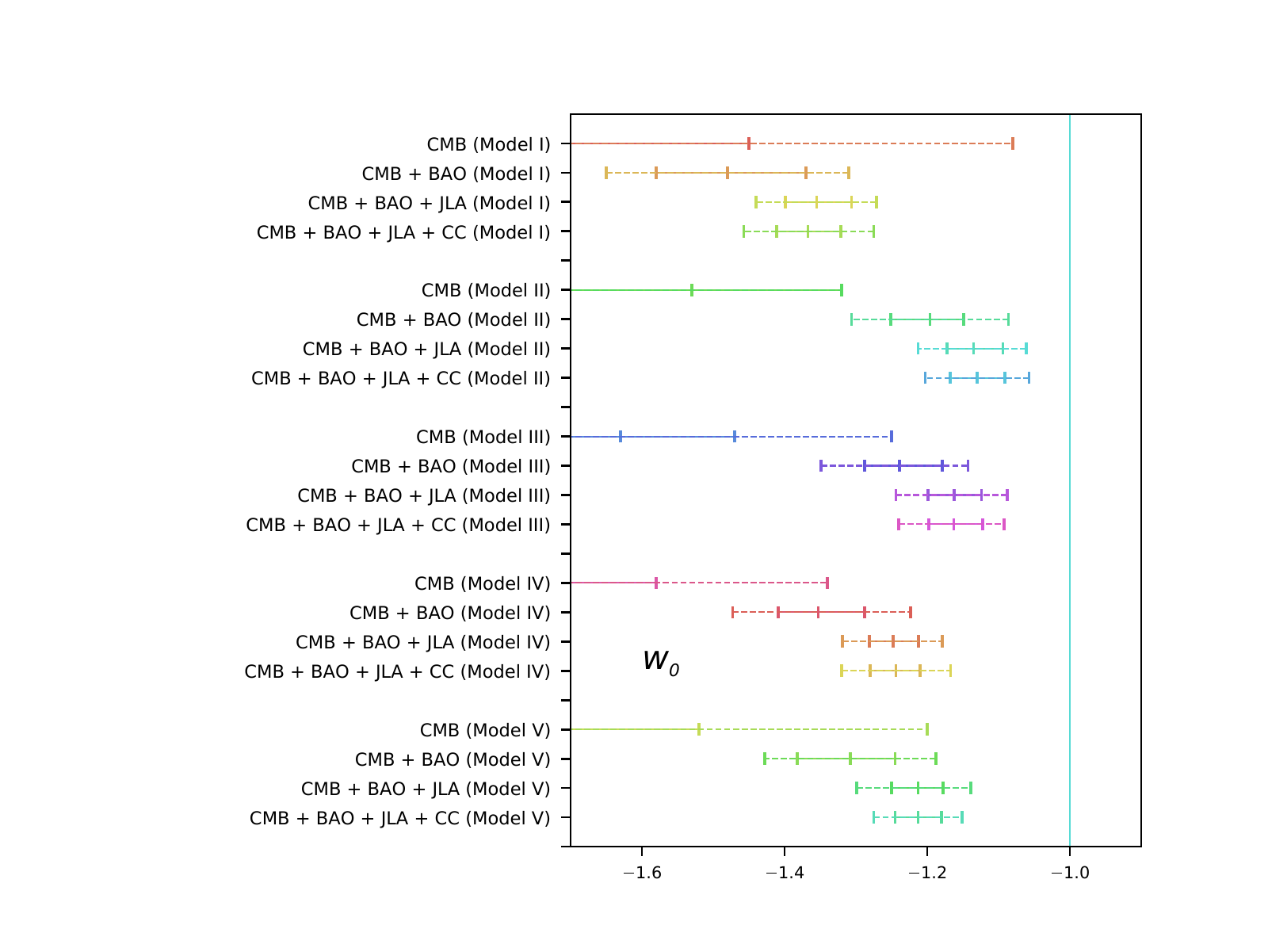}
\caption{{\it{Whisker graph with the 68\% CL (solid line) and 95\% CL (dashed line) regions for the free model parameter $w_0$ of the DE parametrizations  (\ref{model1})-(\ref{model5}), for the combination of datasets considered in this work.}}}
\label{whisker}
\end{figure}

In this section we present the observational constraints and their implications for all 
the one-parameter  DE parametrizations  (\ref{model1})-(\ref{model5}). All  models 
are confronted initially with CMB data alone, and then with different combinations of 
cosmological data. In the Appendix  we  show all  Tables containing the constraints on the model parameters 
for all  observational datasets used in this work. Additionally, in Fig. \ref{whisker} we   concisely display the constraints on the present 
value of the dark energy equation of state $w_0$, for all  models, considering all  observational datasets. 
In the following  we investigate the one parameter DE models in detail, presenting their observational consequences.\\ 

\begin{figure}[ht]
    \centering
    \includegraphics[width=0.49\textwidth]{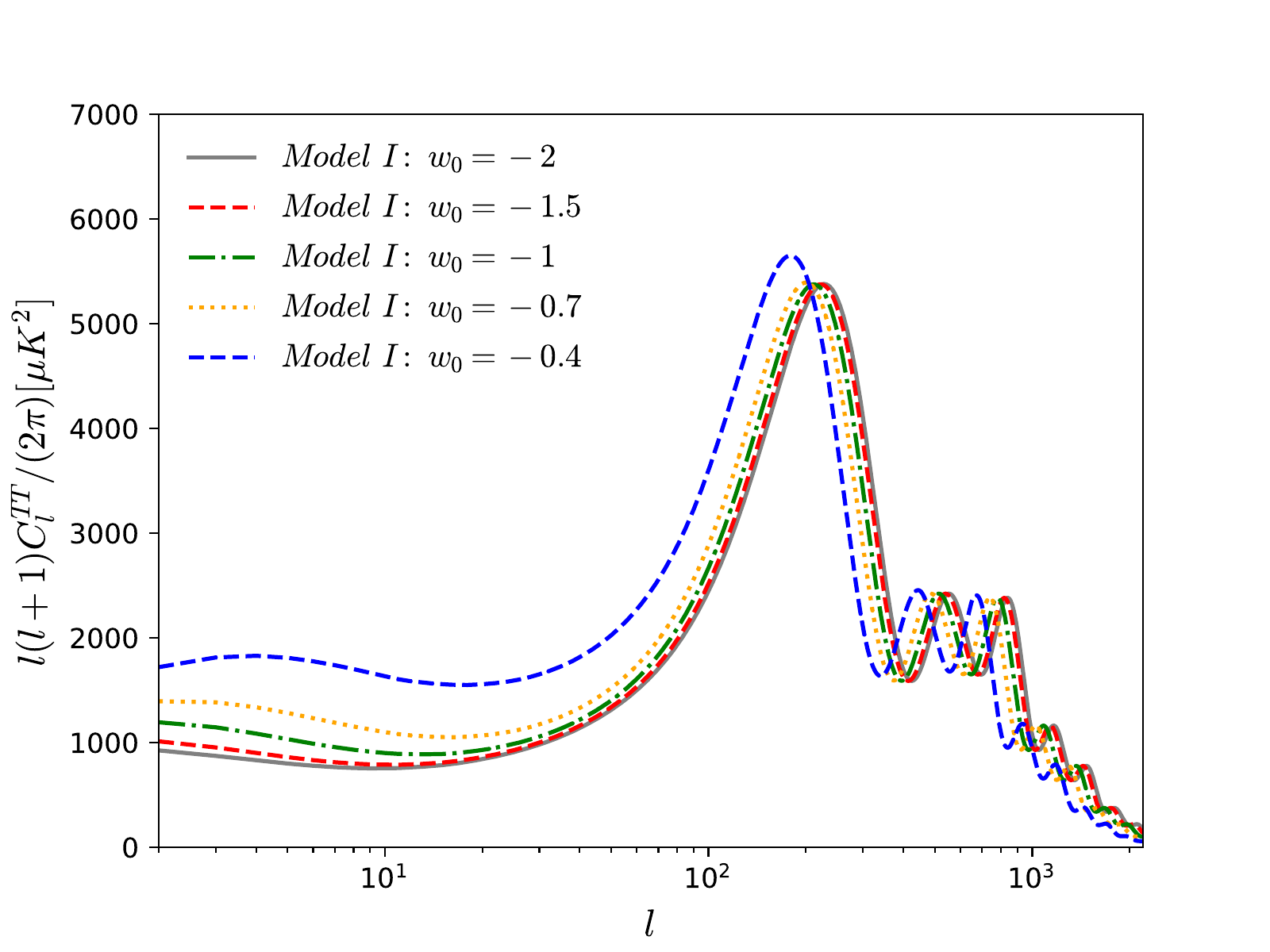}
    \includegraphics[width=0.49\textwidth]{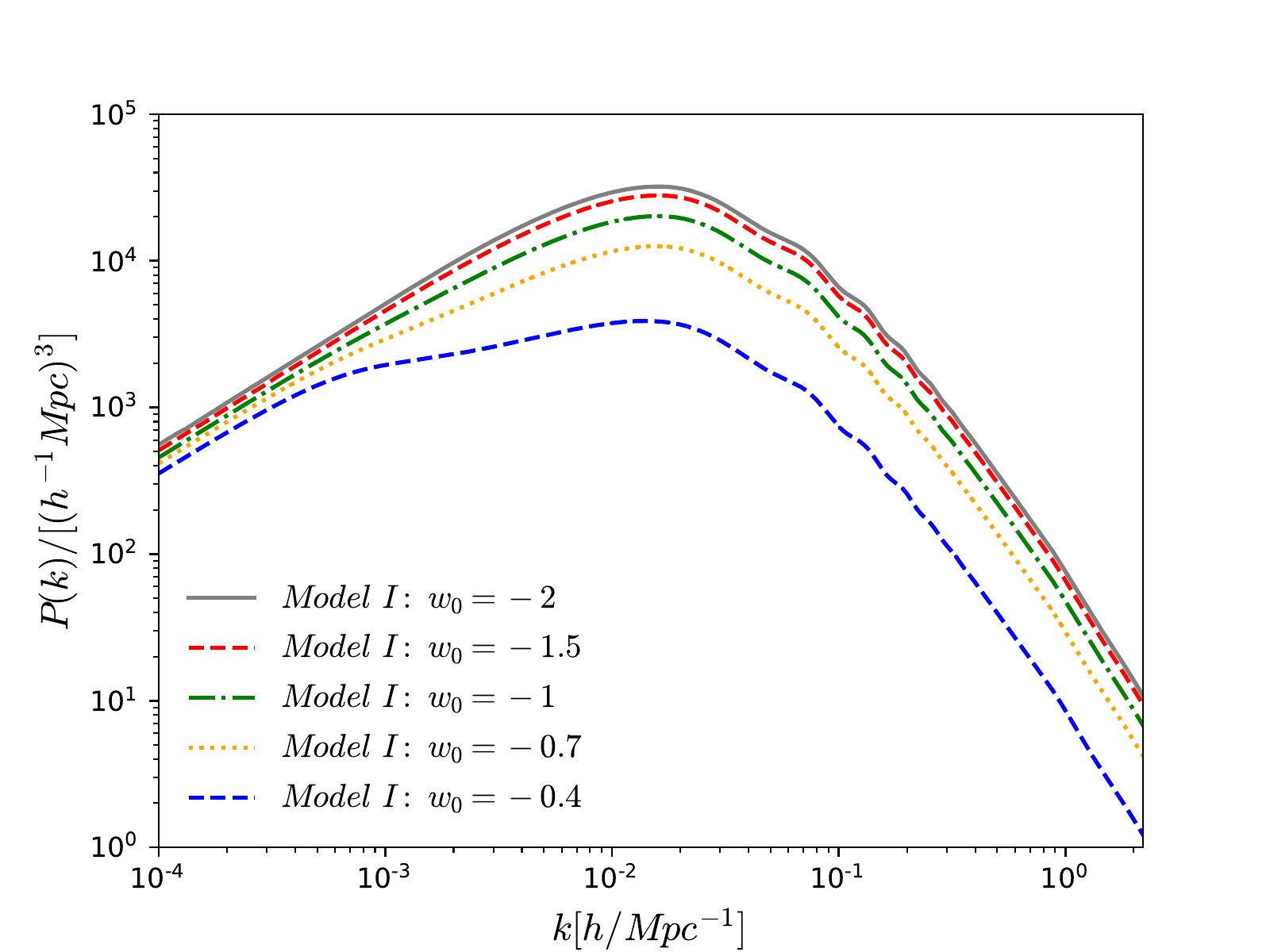}
    \caption{{\it{The  CMB TT spectra (left graph) and the matter power 
spectra (right graph), for  Model I of (\ref{model1}), namely $w_x(a)=w_0\exp(a-1)$, for 
various values of the free model parameter $w_0$.}}}
    \label{fig:model1-cmb-mpower}
\end{figure}

We start by investigating Model I of (\ref{model1}), namely $w_x(a)=w_0\exp(a-1)$. In 
Fig.~\ref{fig:model1-cmb-mpower} we can see the effects of different $w_0$ values on the 
temperature and matter power spectra. The results of the observational analysis of this 
model can be seen in Table \ref{tab:results-model1} of the  Appendix, where we display the 68\% and 95\% 
confidence level (CL) constraints for various quantities 
while the full contour plots are  presented in Fig.~\ref{fig:tri1}.

As we observe from Table \ref{tab:results-model1} (see  Appendix), 
the CMB data alone allow a very large value of the 
Hubble constant at present and moreover its error bars are significantly large: $H_0= 74_{- 7}^{+ 
11}$ at 68\% CL ($H_0 = 74_{-15}^{+14}$ at 95\% CL). The constraint on $H_0$ is actually very close to its local measurements  \cite{Riess:2016jrr}, recently 
confirmed by \cite{R18} and  \cite{Birrer:2018vtm}.
\begin{figure}[ht]
\centering
\includegraphics[width=0.83\textwidth]{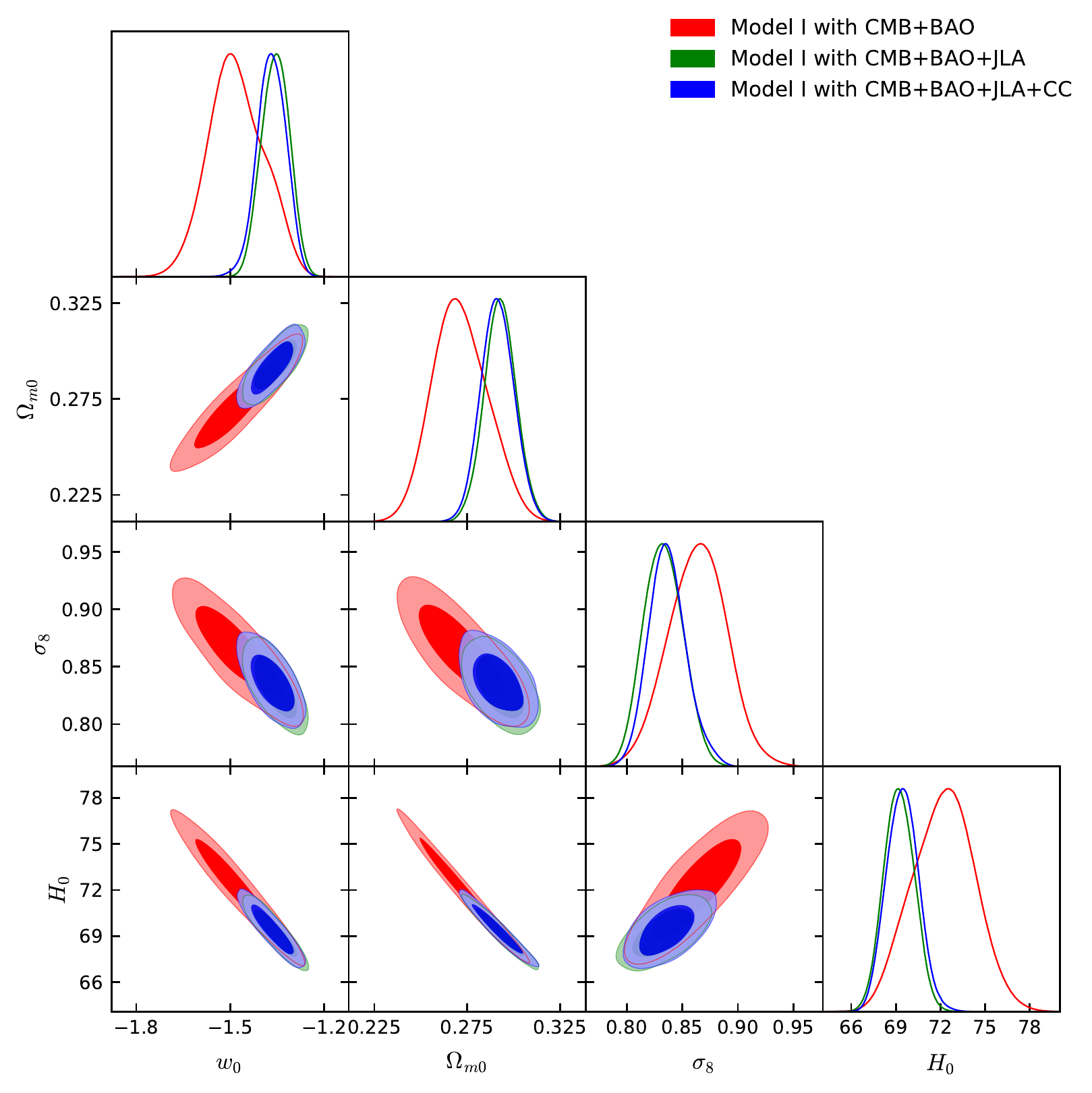}
\caption{{\it{The 2D contour plots for several combinations of various quantities for Model I of (\ref{model1}), namely $w_x(a)=w_0\exp(a-1)$, and the corresponding 1D posterior distributions.}}}
\label{fig:tri1}
\end{figure}

From the  Table \ref{tab:results-model1}   we see that the present value of the DE  equation-of-state parameter for CMB alone is found to prefer a phantom dark energy scenario, namely $w_0 < -1$, at more than 95\% CL.
Consequently, the matter density parameter decreases and acquires a very low value   
($\Omega_{m0}= 0.268_{-0.081}^{+0.038}$ at 68\%  CL). However, since these $\Omega_{m0}$ 
and $\sigma_8$ are anti-correlated, while $\sigma_8$ is positively correlated with $H_0$ 
(see Fig.~\ref{fig:tri1}), hence, this does not correspond to the 
alleviation of the 
$S_8=\sigma_8 \sqrt{\Omega_{m0}/0.3}$ 
tension of Planck's  indirect estimation  with its direct measurements from cosmic shear 
experiments like the Canada France Hawaii Lensing 
Survey (CFHTLenS)~\cite{Heymans:2012gg, Erben:2012zw}, the Kilo Degree Survey of$~$450 
deg$^2$ of imaging data (KiDS-450)~\cite{Hildebrandt:2016iqg}, and the Dark Energy Survey 
(DES)~\cite{Abbott:2017wau}.

When the BAO data are added to CMB, the constraints on the model parameters are 
significantly improved and the error bars on most of the parameters, in particular   
$w_0$, $\Omega_{m0}$, $\sigma_8$ and $H_0$, are decreased. The mean value of the Hubble 
constant slightly shifts towards a lower value, and the DE equation of state at 
present, $w_0$, moves towards a smaller one ($w_0 = -1.48\pm 0.10$ at 68\% CL) comparing 
to its estimation from CMB alone ($w_0 < -1.45$ at 68\% CL). As one can see, the  
CMB+BAO data also assure the validity of $w_0< -1$ at more than 
99\% CL. The interesting output of this analysis is that the constraint on $H_0$ is again 
found to be very close to its estimation by local measurements \cite{Riess:2016jrr}. 

The addition of JLA to the former data set combination (i.e., CMB+BAO) further improves the cosmological 
constraints, as one can clearly see from Table \ref{tab:results-model1} (see Appendix). In particular, 
we see that $H_0$ again shifts down and $w_0$ up, with decreasing error bars. An 
analogous improvement of the bounds can be seen for $\Omega_{m0}$ 
and $\sigma_8$. Although the estimation of $H_0$ from this analysis decreases in 
comparison to the previous results of CMB and CMB+BAO,   within 95\% CL, it can 
still match the direct estimation  \cite{Riess:2016jrr}. Furthermore, the DE 
equation of state at present is again found to be in phantom regime.

We close the analysis by adding the CC dataset, nevertheless the results for the 
data combination CMB+BAO+JLA+CC do not exhibit significant differences from the previous 
case CMB+BAO+JLA.

\begin{figure}[ht]
\centering
\includegraphics[width=0.495\textwidth]{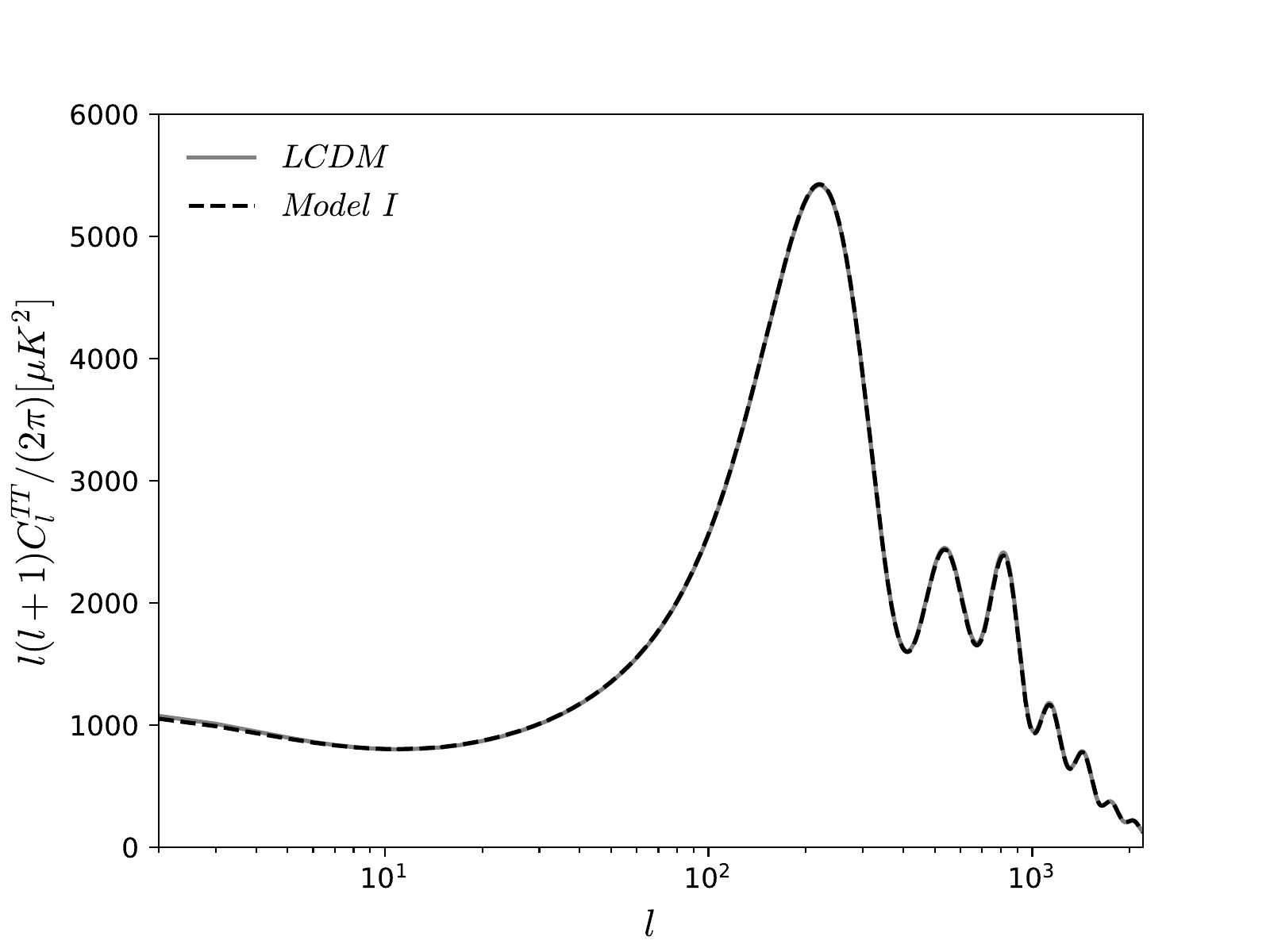}
\includegraphics[width=0.495\textwidth]{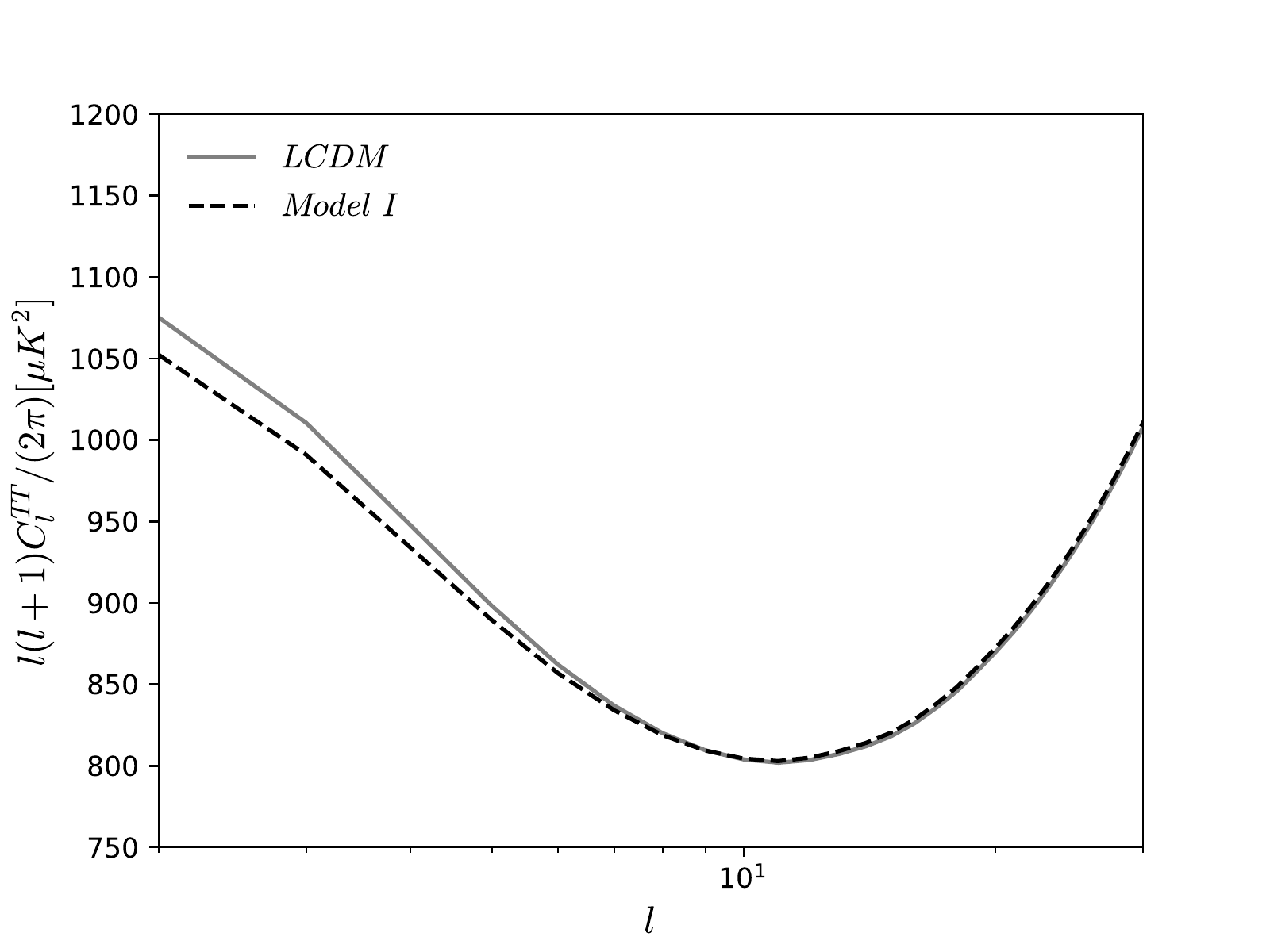}
\caption{{\it{Comparison between the best fit for the $\Lambda$CDM paradigm and for the 
Model I of (\ref{model1}), namely $w_x(a)=w_0\exp(a-1)$. While the curves are almost 
indistinguishable in the high multipole range, at large scales Model I can 
better recover the lower quadrupole of the data.}}}
\label{fig:bf1}
\end{figure}

In summary, the observational analysis for Model I shows that $w_0< -1$ at more than 95\% 
CL for CMB only, while the tension on $H_0$ seems to be alleviated. The addition of JLA 
shifts $H_0$ towards lower values, but still in agreement within 2$\sigma$ with 
\cite{Riess:2016jrr}, while the addition of CC does not affect  the results significantly. The contour plot in the $w_0-H_0$ plane can be seen in lower left graph of Fig.~\ref{fig:tri1}. 
The preference for a phantom DE equation of state is due to the better fit of the large 
scales of the temperature power spectrum, that prefers a lower quadrupole with respect to 
the $\Lambda$CDM scenario, as can be clearly seen in Fig.~\ref{fig:bf1}. 

Finally,  comparing the results obtained for Model I with the constraints 
released by the Planck collaboration \cite{Aghanim:2018eyx} for the $w$CDM or $w_0w_a$CDM 
models, we can notice that for Model I using only CMB data the $H_0$ value is well 
constrained 
by the data and is close to its directly measured value \cite{Riess:2016jrr}, while it 
has a slightly lower limit for the Planck's extended scenario (that means the $w_0w_a$CDM scenario). On the other hand, when 
adding the BAO data, the $H_0$ value is still high and in agreement
with \cite{Riess:2016jrr}, while in the Planck case the Hubble constant decreases leading  to the aforementioned tension. In this context we refer to a recent work \cite{Lin:2017bhs} discussing the tensions in the cosmological parameters from various observational data and some possible explanations. \\


We now investigate Model II of (\ref{model2}), namely  $w_x(a)=w_0a[1-\log(a)]$. 
In Table \ref{tab:results-model2} of the  Appendix we summarize the observational 
constraints arising from various data combinations.  We do not explicitly present the corresponding contour plots since they are similar to the ones of Model I.

Comparing the results with those of Model I above, we can see that for Model II considering the CMB data alone, $H_0$ acquires higher values ($H_0 = 81_{- 9} ^{+   12}$ at 68\% 
CL) and $w_0$ indicates a strong evidence for a phantom equation of state which remains 
at more than 95\% CL. Moreover, similarly to Model I, for Model II we also observe that the combinations CMB+BAO, CMB+BAO+JLA and CMB+BAO+JLA+CC significantly improve the 
constraints and reduce the error bars on the parameters. In particular, the $H_0$ mean value shifts 
towards lower values and we find $w_0< -1$ at more than 95\% CL for all data  combinations. 
Note that from Table \ref{tab:results-model2} (see Appendix) we see that for the data set CMB+BAO the 
estimated value of $H_0$ is in agreement within 1 standard deviation with the local 
estimation of \cite{Riess:2016jrr} and thus the $H_0$-tension is alleviated ($H_0$ is 
higher than the one estimated by Planck 2015 
\cite{Ade:2015xua} for the base $\Lambda$CDM scenario, and it is in perfect agreement to  
 \cite{Riess:2016jrr}). Additionally, for the last two combinations CMB+BAO+JLA and CMB+BAO+JLA+CC we observe that while the Hubble constant is always in 
agreement within $2\sigma$ with \cite{Riess:2016jrr}, in contrast to Model I $w_0$ 
prefers 
a lower phantom mean value, but still with high significance. 
Finally, similar  to the previous model, the  $\sigma_8$ tension is not reconciled. \\


We proceed to the investigation of Model III of (\ref{model3}), namely 
$w_x(a)=w_0a\exp(1-a)$. 
Using the same observational datasets, in Table 
\ref{tab:results-model3} of the Appendix we summarize the observational 
constraints on this model.

As we can see, for the CMB data alone the Hubble parameter acquires an even larger mean 
value in comparison to the previous models, while the DE equation-of-state parameter at 
present obtains a smaller value, namely  $w_0 = -1.63_{-0.37}^{+0.38}$ at 95\% CL. 
Similarly to the previous model, we find that the inclusion of any external data 
set, namely BAO, JLA or CC, to CMB significantly improves the constraints, and  
$w_0 <-1$ is still valid up to 95\% CL. For the combination of CMB+BAO data we see 
that the estimated value of $H_0 =  71.4_{-    1.6}^{+    1.4}$ (at 68\% CL) is 
perfectly in agreement to its local estimation of \cite{Riess:2016jrr}, alleviating the 
$H_0$-tension. Moreover, concerning $w_0$ we can note that it is constrained to be $w_0 = 
-1.239_{- 0.049}^{+    0.060}$ (at 68\% CL) which is phantom at more than $3\sigma$. 
The addition of JLA to CMB+BAO decreases the error bars on  $H_0$, $\Omega_{m0}$ and 
$\sigma_8$, while within 95\% CL this model seems to alleviate the tension on $H_0$. 
Finally, the combination CMB+BAO+JLA+CC does not offer any notable
differences compared to the analysis with CMB+BAO+JLA, and thus similar conclusions are  achieved.\\



We investigate Model IV of (\ref{model4}), namely 
$w_x(a)=w_0a[1+\sin(1-a)]$. 
In Table 
\ref{tab:results-model4} of the Appendix we summarize the observational constraints arising from various 
data combinations.

As we see, the estimations of the Hubble parameter for all the dataset combinations are 
shifted towards higher values than $\Lambda$CDM. For the CMB data only, $H_0$ acquires 
values comparable with Model III, i.e. $H_0 =   84.3_{-    6.5}^{+    9.9}$ at 
68\% CL. 
As before, the inclusion of any external data set significantly improves the constraints  
on the cosmological parameters, decreasing the error bars. A common feature for all the 
analyses is that $w_0$ remains in the phantom regime at more than 95\% CL. Furthermore, 
for this model the CMB+BAO+JLA and CMB+BAO+JLA+CC data combinations favor a phantom DE 
equation of state at many standard deviations. Additionally, it is clearly seen that the 
$H_0$-tension is alleviated for all the combinations considered, apart from the CMB 
data alone which predict a quite high $H_0$ value. \\


We close our analysis with the investigation of  Model V of (\ref{model5}), namely 
$w_x(a)=w_0a[1+\arcsin(1-a)]$. 
In a similar fashion, using the same observational datasets, in Table \ref{tab:results-model5} of the Appendix we summarize the observational 
constraints on this model.

As we observe, we can clearly notice that this model maintains a similar trend compared 
to the previous four dynamical DE models. The present value of the DE equation-of-state 
parameter is constrained in the phantom regime up to 99\% CL. The Hubble parameter acquires a very large value for the CMB data only ($H_0 = 82.9_{-7.0}^{+ 12}$ at 68\% CL) with large error bars, however for the other data combinations $H_0$ and its errors bars decrease, and it  becomes clear that the $H_0$-tension can be alleviated.

\section{Model Comparison and the Bayesian Evidence}
\label{sec-baysian}

\begin{figure}[ht]
\includegraphics[width=0.395\textwidth]{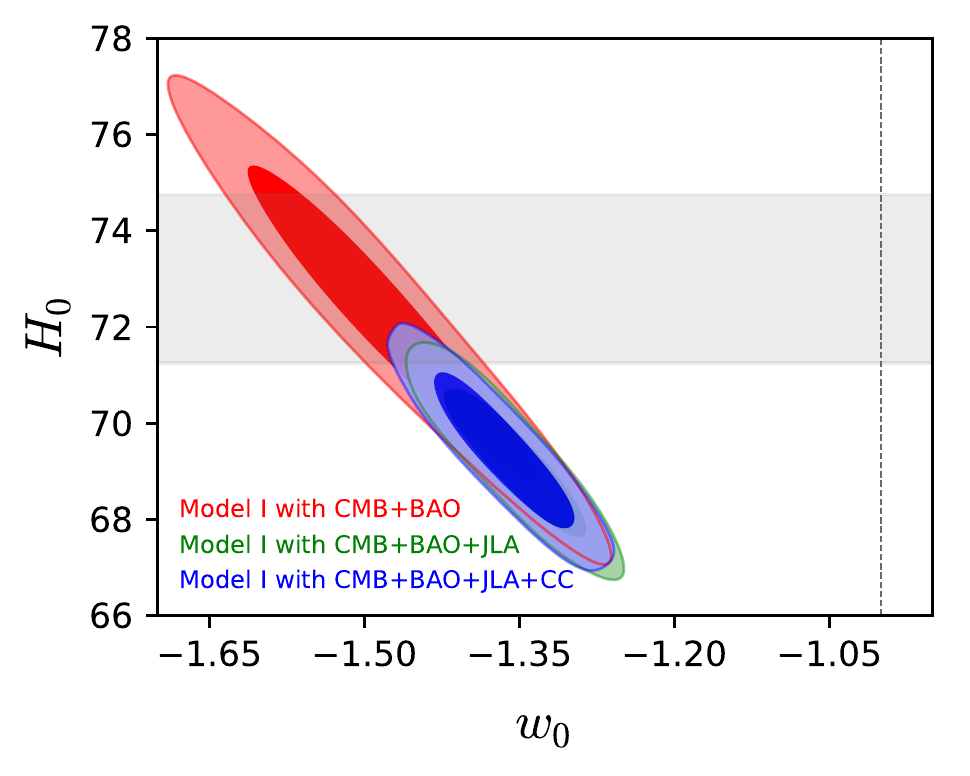}
\includegraphics[width=0.42\textwidth]{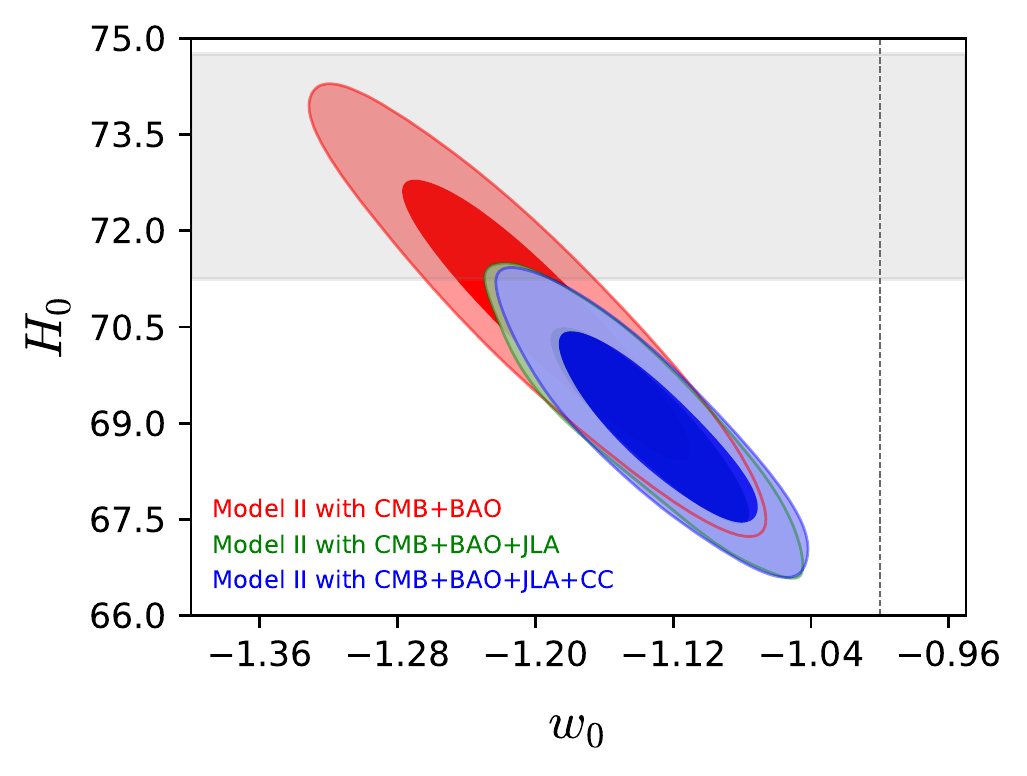}
\includegraphics[width=0.42\textwidth]{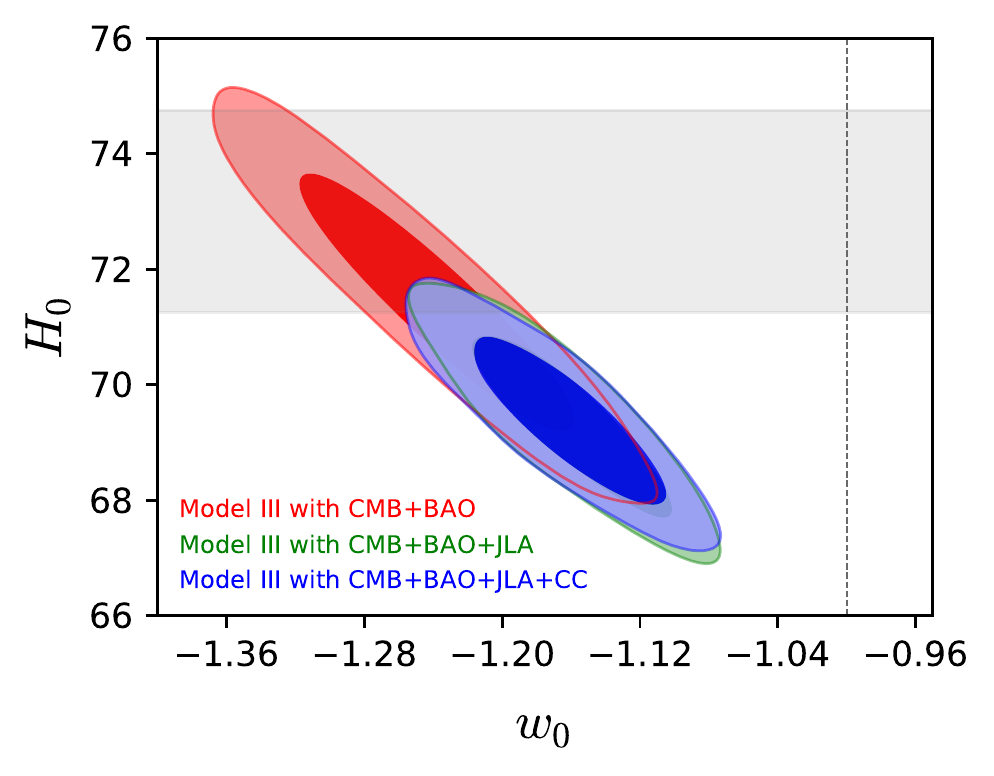}
\includegraphics[width=0.405\textwidth]{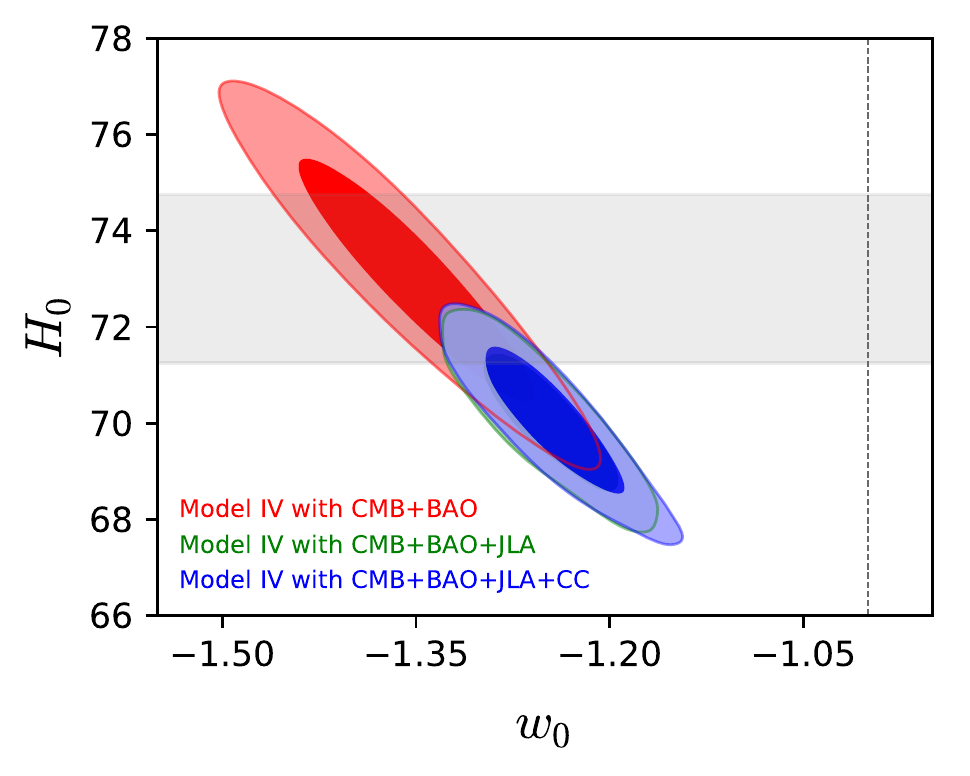}
\includegraphics[width=0.42\textwidth]{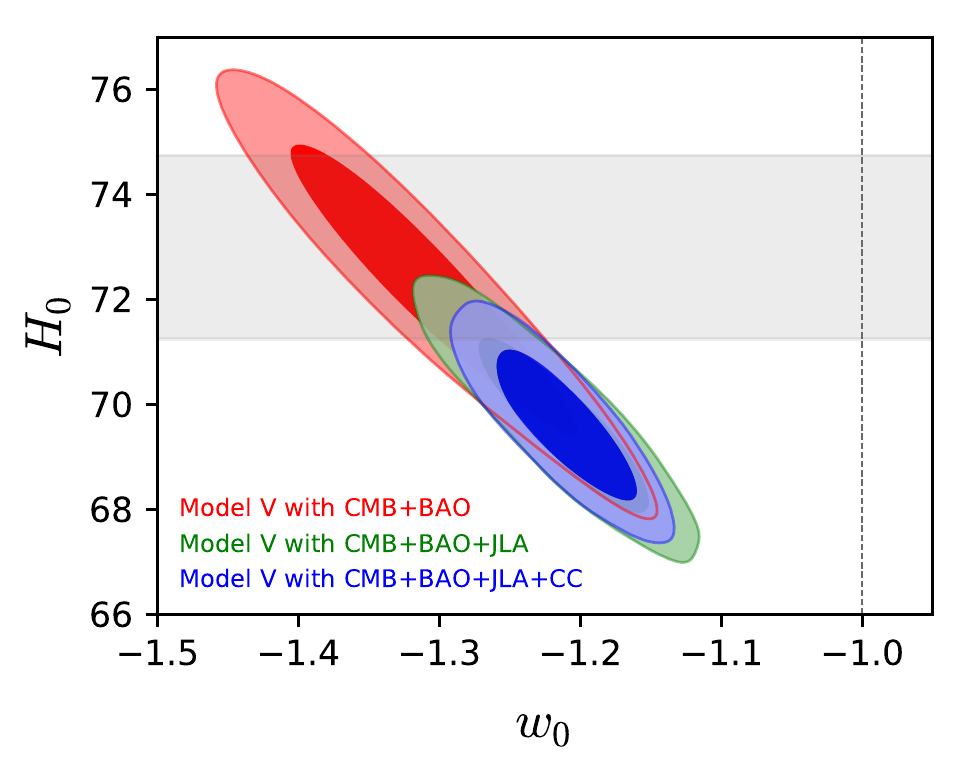}
\caption{{\it{Contour plots at the $68 \%$ and $95 \%$ CL on the $w_0-H_0$
plane for the five models of one-parameter DE parametrizations 
(\ref{model1})-(\ref{model5}), for various combinations of datasets. The gray horizontal 
band is the 68\% CL Hubble parameter value corresponding to the direct measurement of 
\cite{Riess:2016jrr}, while the dotted vertical line marks the cosmological constant 
value $w_0=-1$. 
}}}
\label{H0w0}
\end{figure}

In the previous section we present the observational analysis and we extracted the  
constraints on various cosmological parameters of the five examined models. Concerning  the dark energy equation of state at present, $w_0$, we have already presented its constraints at 68\% and 95\% CL through the Whisker graph
in Fig.~\ref{whisker} for all the one parameter DE models as well as considering all the observational datasets employed in this work. As we mentioned above, we observe that in all models a phantom DE equation-of-state 
parameter at current time is favored.  Moreover, from Fig.~\ref{whisker} we may 
also note that the   extracted $w_0$ for Model II and III using the common datasets, 
namely, CMB+BAO, CMB+BAO+JLA and CMB+BAO+JLA+CC, are relatively close to the cosmological 
constant boundary $w_ 0 = -1$, compared to other three  models.
Additionally, in order to present in a more transparent way the alleviation of the 
$H_0$-tension, in  Fig.~\ref{H0w0} we summarize
 the contour plots in the $w_0-H_0$ plane for all the examined models.  From the figure 
one can notice that the parameters $H_0$ and $w_0$ are correlated to each other.

The question that arises naturally is which of the five models exhibits a better behavior, and moreover how efficient are they comparing to standard  $\Lambda$CDM cosmology. Hence, we close our work with examining the Bayesian evidence of each of the five models analyzed above, compared to the reference $\Lambda$CDM cosmological scenario. The Bayesian evidence plays a crucial role in determining the observational support of any cosmological model. The involved calculation is performed through the publicly available code \texttt{MCEvidence} 
\cite{Heavens:2017hkr,Heavens:2017afc}\footnote{See
\href{https://github.com/yabebalFantaye/MCEvidence}{github.com/yabebalFantaye/MCEvidence}.}. We mention that \texttt{MCEvidence}  needs only the MCMC chains that are used to extract the parameters of the models. 

In Bayesian analysis one needs to evaluate the posterior probability of the model 
parameters $\theta$, given a particular observational dataset $x$ with any prior information for a model $M$. Using the Bayes theorem one can write 
\begin{eqnarray}\label{BE}
p(\theta|x, M) = \frac{p(x|\theta, M)\,\pi(\theta|M)}{p(x|M)},
\end{eqnarray}
where the quantity $p(x|\theta, M)$ refers to the likelihood as a function of $\theta$ 
with $\pi(\theta|M)$ the prior information. The quantity $p(x|M)$ that appears in the 
denominator of (\ref{BE}) is known as the Bayesian evidence used for the model 
comparison. Let us note that this Bayesian evidence is the integral 
over the non-normalized posterior $\tilde{p} (\theta|x, M) \equiv 
p(x|\theta,M)\,\pi(\theta|M)$, given by 
\begin{eqnarray}
E \equiv p(x|M) = \int d\theta\, p(x|\theta,M)\,\pi(\theta|M).
\end{eqnarray} 
 Now, for any cosmological model $M_i$  and the reference model $M_j$ (the reference 
model is the one with respect to which we compare the observational viability),  
the posterior probability is given by the following law:
\begin{eqnarray}
\frac{p(M_i|x)}{p(M_j|x)} = \frac{\pi(M_i)}{\pi(M_j)}\,\frac{p(x| M_i)}{p(x|M_j)} = 
\frac{\pi(M_i)}{
\pi(M_j)}\, B_{ij},
\end{eqnarray}
where the quantity $B_{ij} = \frac{p(x| M_i)}{p(x|M_j)}$ is the Bayes factor of the 
model $M_i$ with respect to the reference model $M_j$. 
Depending on different values of $B_{ij}$ (or equivalently $\ln B_{ij}$) we quantify the 
observational support of the model $M_i$ over the model $M_j$. Here we  use the widely 
accepted Jeffreys scales \cite{Kass:1995loi} shown in Table \ref{tab:jeffreys}, which 
imply that for $B_{ij} > 1 $ the observational data support model $M_i$ more strongly 
than model $M_j$. The negative values of $\ln B_{ij}$ reverse the conclusion, that is the reference model $M_j$ is preferred over $M_i$.  
\begin{center}                 
\begin{table}[!h]                                     
\begin{tabular}{cc}                 
\hline\hline               
$\ln B_{ij}$ &$\ \ $ Strength of evidence for model ${M}_i$ $\ \ $ \\ \hline
$0 \leq \ln B_{ij} < 1$ & Weak \\
$1 \leq \ln B_{ij} < 3$ & Definite/Positive \\
$3 \leq \ln B_{ij} < 5$ & Strong \\
$\ln B_{ij} \geq 5$     & Very strong \\
\hline\hline              
\end{tabular}             
\caption{ Revised Jeffreys scale used to quantify the observational
support of  model $M_i$ with respect to the reference model $M_j$  \cite{Kass:1995loi}. }
\label{tab:jeffreys}         
\end{table}                
\end{center}    
\begin{table}      
\begin{center}                    
\begin{tabular}{ccccccccc}                                      \hline\hline               
Dataset & Model & $\ln B_{ij}$ & ~~Strength of evidence for reference $\Lambda$CDM 
scenario \\ 
\hline
CMB & Model I & $-1.9$ & Definite/Positive\\
CMB+BAO & Model I & $-2.9$ & Definite/Positive\\
CMB+BAO+JLA & Model I & $-6.0$ & Very Strong\\
CMB+BAO+JLA+CC & Model I & $-5.2$ & Very Strong\\
\hline\hline 
CMB & Model II & $-1.7$ & Definite/Positive\\
CMB+BAO & Model II & $-2.3$ & Definite/Positive\\
CMB+BAO+JLA & Model II & $-3.1$ & Strong\\
CMB+BAO+JLA+CC & Model II & $-3.0$ & Strong\\ 
\hline \hline
CMB & Model III & $-1.6$ & Definite/Positive\\
CMB+BAO & Model III & $-2.6$ & Definite/Positive\\
CMB+BAO+JLA & Model III & $-4.2$ & Strong\\
CMB+BAO+JLA+CC & Model III & $-3.9$ & Strong\\ 
\hline \hline 
CMB & Model IV & $-2.1$ & Definite/Positive\\
CMB+BAO & Model IV & $-3.7$ & Strong \\
CMB+BAO+JLA & Model IV & $-6.9$ & Very Strong \\
CMB+BAO+JLA+CC & Model IV & $-7.2$ & Very Strong \\
\hline \hline 
CMB & Model V & $-2.8$ & Definite/Positive \\
CMB+BAO & Model V & $-2.9$ & Definite/Positive\\
CMB+BAO+JLA & Model V & $-5.8$ & Very Strong\\
CMB+BAO+JLA+CC & Model V & $-5.3$ & Very Strong\\
\hline 
\hline\hline
\end{tabular}    
\caption{Summary of the Bayes factors values $\ln B_{ij}$ calculated for the five one-parameter DE models  (\ref{model1})-(\ref{model5}), with respect to the reference 
$\Lambda$CDM scenario. The negative sign indicates that the reference  scenario is preferred over the fitted models.} 
\label{tab:bayesian}                          
\end{center}    
\end{table}        

In Table \ref{tab:bayesian} we present the values of $\ln B_{ij}$ calculated for the five one-parameter DE models  (\ref{model1})-(\ref{model5}) analyzed in the previous 
section, for various observational datasets, compared to the reference $\Lambda$CDM scenario. From the values of $\ln B_{ij}$ we can see that Model II and Model III present 
a better behavior than the other three analyzed models. However, comparing to all models 
the reference $\Lambda$CDM scenario is favored. Nevertheless, we mention here that, interestingly enough, the one-parameter DE parametrizations considered in the present 
work seem to behave similarly or be less disfavored with respect to $\Lambda$CDM  scenario comparing with two-parameter DE parametrizations 
\cite{Ma:2011nc,Feng:2011zzo,Yang:2017alx,Pan:2017zoh}. 
This is an indication that  one-parameter DE models can indeed be 
efficient in describing the universe evolution.

\section{Summary and Conclusions}
\label{sec-conclu}

The phenomenological parametrizations of DE equation of state can be very helpful for the 
investigation of DE features, since they are of general validity and can describe the DE 
sector independently of whether it is an extra peculiar fluid in the framework of general 
relativity or it is effectively of gravitational origin. However, although in the 
literature there has been a large amount of research on DE parametrizations which involve 
two or more free parameters, the one-parameter parametrizations seem to be  underestimated.

In this work we performed a detailed observational confrontation of several one-parameter 
DE parametrizations, with various combination datasets. In particular, we used data from  
 cosmic microwave background (CMB) observations, from Joint light-curve analysis sample 
from Supernovae Type Ia observations (JLA), from baryon acoustic oscillations  (BAO) 
distance measurements, as well as from cosmic chronometers  Hubble parameter measurements 
(CC), and we  additionally performed various combined analyses.

Our analyses revealed that all the examined one-parameter dynamical DE models favor a phantom DE
equation-of-state at present time $w_0$, and this remains valid at more than 95\% CL, confirming the result obtained in various other works in different contexts ~\cite{DiValentino:2017iww,DiValentino:2015ola,DiValentino:2016hlg,DiValentino:2017zyq,Yang:2017ccc,Yang:2017zjs,Mortsell:2018mfj,Yang:2018euj,Yang:2018uae}. The 
inclusion of any external dataset to CMB improves the fitting and 
decreases the errors significantly without any change in the conclusion. Concerning the 
present value of the Hubble parameter $H_0$, we found that the CMB data alone leads to 
large error bars, however the inclusion of other datasets decreases them significantly, 
with the favored $H_0$ value being in perfect agreement with 
its direct measurements. Hence, we deduce that one-parameter DE models can provide a 
solution to the known $H_0$-tension between local measurements and Planck indirect ones. 
This is one of the main results of the present work. Nevertheless, the possible  
$\sigma_8$-tension does not seem to be reconciled, since in all models the favored 
$\sigma_8$ value is similar to the Planck's estimated one.

Lastly, in order to examine which of the five models is better fitted to the data, as 
well as in order to compare them with the standard $\Lambda$CDM cosmological scenario, we compute  their Bayesian evidences using the \texttt{MCEvidence} (summarized in Table \ref{tab:bayesian}). 
As we saw Model II and Model III are relatively close to $\Lambda$CDM (this can also be viewed from the Whisker graph in Fig.~\ref{whisker} where $w_0$ for Model II and Model III are relatively close to $w_0 = -1$ compared to other models). However, the reference 
$\Lambda$CDM scenario is still favored compared to all one parameter dynamical DE models. Nevertheless, these 
one-parameter DE models have similar or better efficiency in fitting the data comparing 
with the two-parameter DE parametrizations analyzed in the literature, taking into 
account their advantage that they are more economical and have one free parameter less. 
This is an indication that one-parameter DE models can indeed be efficient in describing 
the universe evolution, and thus they deserve a thorough investigation.

\begin{acknowledgments}
The authors would like to thank an anonymous referee for essential suggestions that
improved the presentation and the quality of the manuscript.
WY was supported by the National Natural Science Foundation of China under Grants No.  
11705079 and No.  11647153. SP acknowledges the research grant under Faculty Research and Professional Development Fund (FRPDF) Scheme of Presidency University, Kolkata, India.  
EDV acknowledges support from the European Research Council 
in the form of a Consolidator Grant with number 681431. SC acknowledges the Mathematical 
Research Impact Centric Support (MATRICS), project reference no. MTR/2017/000407, by the 
Science and Engineering Research Board, Government of India. This article is based upon 
work from CANTATA COST (European Cooperation in Science and Technology) action CA15117, 
EU Framework Programme Horizon 2020.
\end{acknowledgments}

\appendix*
\section{THE TABLES}
\label{appendix}

In this Appendix we present all Tables that display the summary of the 68\% and 95\% CL constraints, 
using various datasets, for all the one-parameters models investigated in the main text.

\begingroup                                                 \squeezetable                                                   \begin{table}                          
\begin{center}           
\resizebox{\columnwidth}{!}{
\begin{tabular}{cccccccc}           
\hline\hline                
Parameters & CMB & CMB+BAO & CMB+BAO+JLA & CMB+BAO+JLA+CC\\ \hline
$\Omega_{\rm c} h^2$ &  $    0.1200_{- 0.0015- 0.0029}^{+    0.0015+    0.0028}$ & $    
0.1182_{-   
 0.0014- 0.0027}^{+    0.0014+    0.0025}$ & $    0.1172_{-    0.0012-    0.0023}^{+  
0.0012+    0.0023}$ & $    0.1174_{-    0.0012-    0.0023}^{+    0.0012+    0.0024}$ \\
$\Omega_{\rm b} h^2$ & $    0.02218_{-    0.00016-    0.00031}^{+    0.00016 +    
0.00032}$  & $    
0.02230_{-    0.00017-    0.00030}^{+    0.00014 +    0.00032}$ & $    0.02237_{-    
0.00015-    0.
00030}^{+    0.00015+    0.00030}$ & $    0.02235_{-    0.00014-    0.00029}^{+    
0.00015+    0.
00028}$ \\
$100\theta_{MC}$ & $    1.04039_{-    0.00033-    0.00063}^{+    0.00033 +    0.00067}$  
& $    1.
04070_{-    0.00035-    0.00067}^{+    0.00035+    0.00067}$ & $    1.04078_{-    
0.00033-    0.00057}^{+    0.00030 +    0.00063}$  & $    1.04075_{-    0.00029-    
0.00061}^{+    
0.00030+    0.00059}$ \\
$\tau$ &   $    0.081_{-    0.017-    0.035}^{+    0.019+    0.033}$ & $    0.091_{-    
0.018-    0.035}^{+    0.018 +    0.033}$ & $    0.098_{-    0.017-    0.033}^{+    
0.017+ 
   0.034}$ 
& $    0.097_{-    0.017-    0.032}^{+    0.017+    0.033}$ \\
$n_s$ & $    0.9727_{-    0.0046-    0.0092}^{+    0.0046+    0.0090}$ & $ 0.9772_{-    
0.0054-    0.0088}^{+    0.0045+    0.0100}$ & $    0.9794_{-    0.0040-    0.0080}^{+    
0.0042+    
0.0081}$ &
 $    0.9789_{-    0.0041-    0.0081}^{+    0.0039+    0.0082}$ \\
${\rm{ln}}(10^{10} A_s)$ &  $    3.106_{-    0.033-    0.068}^{+    0.037+    0.066}$ & $ 
   3.121_{
-    0.034-    0.069}^{+    0.034+    0.070}$ & $    3.133_{-    0.033-    0.065}^{+    
0.032+    0.
065}$ & $    3.132_{-    0.032-    0.063}^{+    0.033+    0.065}$ \\
$w_0$ & $   <-1.45\, <-1.08$ & $   -1.48_{-    0.10-    0.17}^{+    0.11+    0.17}$ & $   
-1.355_{-    0.044-    0.085}^{+    0.049+    0.084}$ & $   -1.367_{-    0.044-    
0.090}^{+    0.048+    0.092}$ \\
$\Omega_{m0}$ & $    0.268_{-    0.081-    0.10}^{+    0.038+    0.13}$ & $    0.271_{-   
 0.017-   
 0.027}^{+    0.014+    0.029}$ & $    0.293_{-    0.008-    0.017}^{+    0.008+    
0.017}$ & $    
0.291_{-    0.009-    0.017}^{+    0.009+    0.018}$ \\
$\sigma_8$ & $    0.887_{-    0.061-    0.14}^{+    0.094+    0.13}$ & $    0.864_{-    
0.025-    0.054}^{+    0.029+    0.051}$ & $    0.833_{-    0.019-    0.032}^{+    0.017+ 0.034}$ & 
$ 0.837_{-    0.018-    0.034}^{+    0.016+    0.036}$ \\
$H_0$ & $   74_{-    7-   15}^{+   11+   14}$ & $   72.2_{-    2.1-    3.9}^{+    2.3+    
3.9}$ & $   69.2_{-    1.0-    1.9}^{+    1.1+    2.0}$ & $   69.5_{-    1.1-    2.2}^{+  1.1+   2.1}$ \\
\hline\hline                           
\end{tabular}  }
\caption{Summary of the 68\% and 95\% CL constraints on   Model I of 
(\ref{model1}), namely $w_x(a)=w_0\exp(a-1)$, using various combinations of the 
observational 
data sets. $\Omega_{m0}$ is the current value of $\Omega_m = \Omega_b 
+\Omega_c$, while $H_0$ is in   units 
of km$\,$s$^{-1}\!$Mpc$^{-1}$.
 }\label{tab:results-model1}                                                              
\end{center}                 
\end{table}                                 
\endgroup    
\begingroup                                                         \squeezetable                                                       \begin{table}                         
\begin{center}            
\resizebox{\columnwidth}{!}{
\begin{tabular}{ccccccccccccccc}                                                 
\hline\hline                               
Parameters & CMB & CMB+BAO & CMB+BAO+JLA & CMB+BAO+JLA+CC\\ \hline
$\Omega_{\rm c} h^2$ & $    0.1198_{-    0.0016-    0.0030}^{+    0.0014+    0.0031}$ &  
$  0.1191_{-    0.0012-    0.0025}^{+    0.0013+    0.0025}$ & $    0.1187_{-  0.0011- 0.0022}^{+    0.0012+    0.0024}$ & $    0.1186_{-    0.0012-    0.0023}^{+    0.0012+  0.0023}$  \\

$\Omega_{\rm b} h^2$ & $    0.02220_{-    0.00016-    0.00031}^{+    0.00016+    
0.00031}$ & $    0.02224_{-    0.00016-    0.00029}^{+    0.00014+ 0.0003}$ & $    0.02226_{-    0.00014- 0.00027}^{+    0.00014+    0.00026}$ & $    0.02227_{-    0.00015-    0.00030}^{+ 0.00014+    0.00029}$ \\

$100\theta_{MC}$ & $    1.04042_{-    0.00035-    0.00069}^{+    0.00035+    0.00068}$ & 
$ 1.04054_{-    0.00031-    0.00062}^{+    0.00034+    0.00058}$ & $    1.04057_{-  0.00030-    0.
00060}^{+    0.00032+    0.00059}$ & $    1.04062_{-    0.00032-    0.00062}^{+ 0.00034+    0.00059}$ \\

$\tau$ & $    0.078_{-    0.018-    0.034}^{+    0.018+    0.035}$ & $    0.088_{-  0.016-    0.034}^{+    0.019+    0.031}$ & $    0.092_{-    0.017-    0.035}^{+    0.017+    0.033}$ 
& $    0.092_{-    0.017-    0.033}^{+    0.018+    0.034}$  \\
$n_s$ & $    0.9728_{-    0.0047-    0.0090}^{+    0.0046+    0.0094}$ & $    0.9752_{-  0.0044-  0.0078}^{+    0.0042+    0.0078}$ & $    0.9766_{-    0.0041-    0.0082}^{+    0.0041+  0.0081} $ & $    0.9769_{-    0.0043-    0.008}^{+    0.0043+    0.0082}$  \\

${\rm{ln}}(10^{10} A_s)$ & $    3.100_{-    0.035-    0.065}^{+    0.035+    0.068}$ & $  3.118_{-    0.031-    0.066}^{+    0.036+    0.060}$  & $    3.125_{-    0.033-   0.068}^{+  0.033+    0.064}$ & $    3.124_{-    0.035-    0.066}^{+    0.034+  0.066}$ \\
$w_0$ & $   -1.53_{-    0.39}^{+    0.21}\, <-1.32$ & $   -1.196_{-    0.055-    0.11}^{+ 0.057+    0.11}$ & $   -1.135_{-    0.037-    0.078}^{+    0.041+    0.074}$ & $   -1.130_{-    0.038- 0.073}^{+    0.039+    0.073}$ \\

$\Omega_{m0}$ & $    0.225_{-    0.074-    0.093}^{+    0.033+    0.120}$ & $    0.284_{- 0.012-    0.022}^{+    0.011+    0.022}$ & $    0.297_{-    0.009-    0.017}^{+  0.009+  0.017}$ & $    0.298_{-    0.009-    0.017}^{+    0.009+    0.017}$ \\

$\sigma_8$ & $    0.95_{-    0.07-    0.16}^{+    0.10+    0.14}$ & $    0.853_{-  0.021-    0.040}^{+    0.020+    0.040}$ & $    0.836_{-    0.018-    0.038}^{+    0.018+ 0.036}$ 
& $  0.833_{-    0.017-    0.034}^{+    0.019+    0.034}$ \\

$H_0$ & $   81_{-    9-   18}^{+   12+   18}$ & $   70.7_{-    1.4-    2.9}^{+    1.4+  2.9}$ & $   69.0_{-    1.1-    1.9}^{+    1.0+    2.0}$ & $   68.9_{-    1.1-    1.8}^{+  0.9+    1.9}$ \\
\hline\hline       
\end{tabular} }        
\caption{
Summary of the 68\% and 95\% CL constraints on Model II of (\ref{model2}), namely 
$w_x(a)=w_0a[1-\log(a)]$, using various combinations of the observational data sets. 
$\Omega_{m0}$ is the current value of $\Omega_m = \Omega_b +\Omega_c$, while $H_0$ is in  
 units of km$\,$s$^{-1}\!$Mpc$^{-1}$. 
}\label{tab:results-model2}                                      
\end{center}                
\end{table}              
\endgroup      
\begingroup             
\squeezetable   
\begin{table}[ht]                  
\begin{center}                                                          
\resizebox{\columnwidth}{!}{ 
\begin{tabular}{ccccccccccc}                                                              
\hline\hline                     
Parameters & CMB & CMB+BAO & CMB+BAO+JLA & CMB+BAO+JLA+CC\\ \hline
$\Omega_{\rm c} h^2$ & $    0.1198_{-    0.0015-    0.0028}^{+    0.0015+    0.0029}$ & $ 
   0.1196_{-    0.0012-    0.0022}^{+    0.0012+    0.0023}$  & $    0.1190_{-    0.0012- 
   0.0022}^{+    0.0011+    0.0023}$ & $    0.1190_{-    0.0013-    0.0024}^{+    0.0012+ 
   0.0024}$ \\
$\Omega_{\rm b} h^2$ & $    0.02220_{-    0.00015-    0.00031}^{+    0.00016+    
0.00031}$ & $    0.02221_{-    0.00014-    0.00026}^{+    0.00014+    0.00027}$ & $    
0.02223_{-    0.00015-    0.
00029}^{+    0.00014+    0.00027}$ & $    0.02223_{-    0.00015-    0.00029}^{+    
0.00015+    0.00030}$ \\
$100\theta_{MC}$ & $    1.04044_{-    0.00033-    0.00068}^{+    0.00034+    0.00069}$ & 
$    1.04045_{-    0.00030-    0.00058}^{+    0.00029+    0.00060}$ & $    1.04052_{-    
0.00033-    0.
00061}^{+    0.00030+    0.00061}$ & $    1.04052_{-    0.00030-    0.00064}^{+    
0.00030+    0.00060}$ \\
$\tau$ & $    0.078_{-    0.017-    0.034}^{+    0.017+    0.032}$ & $    0.089_{-    
0.017-    0.033}^{+    0.017+    0.032}$ & $    0.094_{-    0.017-    0.033}^{+    0.017+ 0.033}$ & $    0.
095_{-    0.018-    0.035}^{+    0.018+    0.033}$ \\
$n_s$ & $    0.9733_{-    0.0045-    0.0088}^{+    0.0046+    0.0088}$ & $    0.9748_{-   
 0.0038-    0.0077}^{+    0.0040+    0.0073}$ & $    0.9767_{-    0.0044-    0.0084}^{+   0.0044+    0.0086}
$ & $    0.9765_{-    0.0044-    0.0083}^{+    0.0042+    0.0088}$ \\
${\rm{ln}}(10^{10} A_s)$ & $    3.100_{-    0.032-    0.066}^{+    0.035+    0.064}$ & $  
  3.119_{-    0.033-    0.066}^{+    0.034+    0.064}$ & $    3.128_{-    0.033-    
0.067}^{+    0.033+    0.
065}$ & $    3.129_{-    0.034-    0.068}^{+    0.035+    0.065}$ \\
$w_0$ & $   -1.63_{-    0.33}^{+    0.16}\,<-1.25$ & $   -1.239_{- 0.049-    0.11}^{+  0.060+    0.096}$ & $   -1.162_{-    0.037-    0.072}^{+    0.038+    0.074}$ & $   
-1.163_{-    0.035-    
0.077}^{+    0.041+    0.071}$ \\
$\Omega_{m0}$ & $    0.207_{-    0.056-    0.070}^{+    0.028+    0.086}$ &  $    
0.279_{-    0.012-    0.021}^{+    0.011+    0.023}$ & $    0.295_{-    0.010-  0.017}^{+    0.009+    
0.017}$ & $  0.295_{-    0.008-    0.016}^{+    0.008+    0.016}$ \\
$\sigma_8$ & $    0.969_{-    0.059-    0.13}^{+    0.085+    0.12}$ & $    0.857_{-0.021-    0.
042}^{+    0.020+    0.042}$ & $    0.832_{-    0.017-   0.036}^{+ 0.017+    0.037}$ & 
$    0.833_{-    0.018-    0.034}^{+    0.018+    0.033}$  \\
$H_0$ & $   84._{-    8-   15}^{+   10+   15}$ & $   71.4_{-    1.6-    2.8}^{+    1.4+ 3.0}$ & $   69.4_{-    1.0-    1.9}^{+    1.0+    2.0}$ & $   69.4_{-    1.0-    1.8}^{+ 0.9+    1.9}$ \\
\hline\hline             
\end{tabular}   }           
\caption{
Summary of the 68\% and 95\% CL constraints on Model III of (\ref{model3}), namely 
$w_x(a)=w_0a\exp(1-a)$, using various combinations of the observational data sets. 
$\Omega_{m0}$ is the current value of $\Omega_m = \Omega_b +\Omega_c$, while $H_0$ is in  
 units of km$\,$s$^{-1}\!$Mpc$^{-1}$.  
}\label{tab:results-model3}            
\end{center}                                
\end{table}                                                                               
\endgroup   

\begingroup               
\squeezetable      
\begin{table}                            
\begin{center}                                                    
\resizebox{\columnwidth}{!}{                            
\begin{tabular}{cccccc}                       
\hline\hline                             
Parameters & CMB & CMB+BAO & CMB+BAO+JLA & CMB+BAO+JLA+CC \\ \hline
$\Omega_{\rm c} h^2$ & $    0.1201_{-    0.0015-    0.0029}^{+    0.0015+    0.0031}$ & $ 
  0.1199_{-    0.0012-    0.0022}^{+    0.0012+    0.0022}$ & $    0.1196_{-    0.0011-   
 0.0022}^{+    0.0011+    0.0022}$ & $    0.1194_{-    0.0011-    0.0021}^{+    0.0010+   0.0021}$\\
$\Omega_{\rm b} h^2$ & $    0.02217_{-    0.00015-    0.00031}^{+    0.00017+    
0.00030}$ & $    0.
02217_{-    0.00015-    0.00029}^{+    0.00015+    0.00029}$
& $    0.02218_{-    0.00014-    0.00029}^{+    0.00014+    0.00026}$ & $    0.02220_{-   
 0.00014-    0.00028}^{+    0.00015+    0.00028}$ \\
$100\theta_{MC}$ & $    1.04038_{-    0.00033-    0.00066}^{+    0.00033+    0.00067}$ & 
$    1.04038_{-    0.00030-    0.00056}^{+    0.00030+    0.00058}$ & $    1.04042_{-    
0.00028-    0.00062}^{+    0.00032+    0.00057}$ & $    1.04044_{-    0.00033-    
0.00059}^{+    
0.00030+    0.00062}$  \\
$\tau$ & $    0.080_{-    0.018-    0.033}^{+    0.017+    0.034}$ & $    0.091_{-    
0.017-    0.034}^{+    0.017+    0.035}$ & $    0.098_{-    0.017-    0.031}^{+    0.017+ 0.032}$ 
& $    0.099_{-    0.017-    0.035}^{+    0.019+    0.033}$ \\
$n_s$ & $    0.9728_{-    0.0043-    0.0088}^{+    0.0043+    0.0086}$ & $    0.9747_{-   
 0.0041-    0.0079}^{+    0.0040+    0.0078}$  & $    0.9765_{-    0.0040-    0.0081}^{+  
  0.0041+    0.0077}
$ &  $    0.9769_{-    0.0040-    0.0078}^{+    0.0041+    0.0079}$  \\
${\rm{ln}}(10^{10} A_s)$ & $    3.104_{-    0.033-    0.064}^{+    0.032+    0.066}$ & $  
  3.125_{-    0.033-    0.067}^{+    0.033+    0.070}$ & $    3.136_{-    0.033-    
0.061}^{+    0.033+    0.
062}$ & $    3.137_{-    0.034-    0.068}^{+    0.037+    0.065}$ \\
$w_0$ & $   <-1.58 \, <-1.34$ & $   -1.353_{-    0.056-    0.12}^{+    0.065+    0.12}$ & 
$   -1.248_{-    0.033-    0.071}^{+    0.036+    0.069}$  &  $   -1.244_{-    0.036-    
0.076}^{+    0.034+    0.077}$\\
$\Omega_{m0}$ & $    0.206_{-    0.052-    0.062}^{+    0.025+    0.080}$ & $    0.268_{- 
  0.012-    0.023}^{+    0.012+    0.024}$ & $    0.290_{-    0.008-    0.016}^{+    
0.008+    0.016}$ & $   
 0.290_{-    0.010-    0.018}^{+    0.008+    0.018}$ \\
$\sigma_8$ & $    0.966_{-    0.050-    0.12}^{+    0.084+    0.11}$ & $    0.865_{-    
0.021-    0.043}^{+    0.021+    0.043}$ & $    0.832_{-    0.018-    0.034}^{+    0.018+ 
   0.036}$ & $    0.830_{-    0.018-    0.036}^{+    0.019+    0.035}$ \\
$H_0$ & $   84.3_{-    6.5-   14}^{+    9.9+   13}$ & $   73.1_{-    1.8-    3.1}^{+    
1.6+    3.3}$ & $   70.07_{-    0.94-    1.9}^{+    0.91+    1.9}$ & $   70.04_{-    
0.97-    2.1}^{+    1.01+    2.1}$\\
\hline\hline              
\end{tabular}   }           
\caption{
Summary of the 68\% and 95\% CL constraints on Model IV of (\ref{model4}), namely 
 $w_x(a)=w_0a[1+\sin(1-a)]$, using various combinations of the observational data sets. 
$\Omega_{m0}$ is the current value of $\Omega_m = \Omega_b +\Omega_c$, while $H_0$ is in  
 units of km$\,$s$^{-1}\!$Mpc$^{-1}$. 
}\label{tab:results-model4}              
\end{center}                           
\end{table}               
\endgroup  

\begingroup              
\squeezetable                 
\begin{table}                        
\begin{center}                        
\resizebox{\columnwidth}{!}{       
\begin{tabular}{cccccc}          
\hline\hline                       
Parameters & CMB & CMB+BAO & CMB+BAO+JLA & CMB+BAO+JLA+CC \\ \hline
$\Omega_{\rm c} h^2$ & $    0.1201_{-    0.0016-    0.0033}^{+    0.0016+    0.0033}$ & $ 
   0.1196_{-    0.0011-    0.0023}^{+    0.0012+    0.0023}$ & $    0.1191_{-    0.0012-  
  0.0024}^{+    0.0012+    0.0023}$ & $    0.1191_{-    0.0013-    0.0023}^{+    0.0011+  
  0.0024}$\\
$\Omega_{\rm b} h^2$ & $    0.02217_{-    0.00017-    0.00034}^{+    0.00017+    
0.00034}$ &  $    
0.02219_{-    0.00014-    0.00029}^{+    0.00014+    0.00027}$ & $    0.02223_{-    
0.00014-    0.00029}^{+    0.00014+    0.00027}$  & $    0.02223_{-    0.00014-    
0.00027}^{+    0.00014+    0.
00028}$ \\
$100\theta_{MC}$ & $    1.04037_{-    0.00034-    0.00069}^{+    0.00035+    0.00067}$ 
& $    1.04042_{-    0.00030-    0.00058}^{+    0.00029+    0.00060}$ &  $    
1.04046763_{-    0.00028-    0.00062}^{+    0.00034+    0.00058}$ & $    1.04053_{-    
0.00031-    0.00059}^{+    0.
00031+    0.00060}$ \\
$\tau$ & $    0.079_{-    0.018-    0.035}^{+    0.018+    0.035}$ & $    0.092_{-    
0.017-    0.036}^{+    0.018+    0.035}$ & $    0.099_{-    0.017-    0.037}^{+    0.017+ 
  0.037}$ & $    0.097_{-    0.017-    0.034}^{+    0.017+    0.034}$ \\
$n_s$ & $    0.9727_{-    0.0046-    0.0091}^{+    0.0045+    0.0090}$ &  $    0.9749_{-  
  0.0042-    0.0080}^{+    0.0041+    0.0086}$& $    0.9772_{-    0.0046-    0.0081}^{+   
 0.0040+    0.0087}
$  & $    0.9766_{-    0.0040-    0.0079}^{+    0.0040+    0.0080}$  \\
${\rm{ln}}(10^{10} A_s)$ & $    3.103_{-    0.035-    0.068}^{+    0.035+    0.067}$ & $  
  3.125_{-    0.034-    0.070}^{+    0.035+    0.065}$ &$    3.136_{-    0.033-    
0.073}^{+    0.033+    0.065}$  & $    3.134_{-    0.033-    0.065}^{+    0.033+    
0.065}$ \\
$w_0$ & $   <-1.52 \, <-1.20$ & $   -1.308_{-    0.074-    0.12}^{+    0.063+    0.12}$ & 
$   -1.213_{-    0.037-    0.086}^{+    0.045+    0.074}$ & $   -1.213_{-    0.032-    
0.062}^{+    0.033+ 
   0.062}$  \\
$\Omega_{m0}$ & $    0.215_{-    0.064-    0.080}^{+    0.027+    0.105}$ & $    0.273_{- 
   0.015-    0.025}^{+    0.012+    0.027}$ & $    0.293_{-    0.009-    0.019}^{+    
0.009+    0.018}$ & $   
 0.293_{-    0.009-    0.017}^{+    0.008+    0.017}$ \\
$\sigma_8$ & $    0.955_{-    0.059-    0.15}^{+    0.094+    0.13}$ & $    0.861_{-    
0.025-    0.045}^{+    0.023+    0.048}$ &  $    0.831_{-    0.017-    0.035}^{+    
0.017+ 
   0.038}$ & $    0.830_{-    0.016-    0.033}^{+    0.016+    0.035}$ \\
$H_0$ & $   82.9_{-    7.0-   17}^{+   12+   15}$ & $   72.3_{-    1.7-    3.4}^{+    
2.0+    3.3}$ & $   69.6_{-    1.2-    2.0}^{+    1.0+    2.3}$ & $   69.64_{-    0.94-   
 1.8}^{+    0.92+    1.9}$ \\
\hline\hline     
\end{tabular}}       
\caption{Summary of the 68\% and 95\% CL constraints on Model V of (\ref{model5}), 
namely $w_x(a)=w_0a[1+\arcsin(1-a)]$, using various combinations of the observational data sets. Here, $\Omega_{m0}$ is the current value of $\Omega_m = \Omega_b +\Omega_c$, while $H_0$ is in  
 units of km$\,$s$^{-1}\!$Mpc$^{-1}$. 
}\label{tab:results-model5}                                                             
 
\end{center}  
\end{table}                                                       
\endgroup


\end{document}